\documentclass{article}%
\usepackage{hyperref}
\usepackage{amsmath}
\usepackage{amsfonts}
\usepackage{amssymb}
\usepackage{graphicx}%
\begin{document}

\title{Finite BRST-antiBRST Transformations in Generalized Hamiltonian Formalism}
\author{\textsc{Pavel Yu. Moshin${}^{a}$\thanks{moshin@rambler.ru \hspace{0.5cm}
${}^{\dagger}$reshet@ispms.tsc.ru}\ \ and Alexander A. Reshetnyak${}%
^{b,c\ddagger}$}\\\textit{${}^{a}$Department of  Physics, Tomsk
State University, 634050, Tomsk,
Russia,}\\\textit{${}^{b}$Institute of Strength Physics and
Materials Science,}\\\textit{Siberian Branch of Russian Academy of
Sciences, 634021, Tomsk, Russia,}\\\textit{${}^{c}$Tomsk State
Pedagogical University, 634061, Tomsk, Russia }} \maketitle

\begin{abstract}
We introduce the notion of finite BRST-antiBRST transformations for
constrained dynamical systems in the generalized Hamiltonian formalism, both
global and field-dependent, with a doublet $\lambda_{a}$, $a=1,2$, of
anticommuting Grassmann parameters and find explicit Jacobians corresponding
to these changes of variables in the path integral. It turns out that the
finite transformations are quadratic in their parameters. Exactly as in the
case of finite field-dependent BRST-antiBRST transformations for the
Yang--Mills vacuum functional in the Lagrangian formalism examined in our
previous paper [arXiv:1405.0790[hep-th]], special field-dependent
BRST-antiBRST transformations with functionally-dependent parameters
$\lambda_{a}=\int dt\ \left(  s_{a}\Lambda\right)  $, generated by a finite
even-valued function $\Lambda\left(  t\right)  $ and by the anticommuting
generators $s_{a}$ of BRST-antiBRST transformations, amount to a precise
change of the gauge-fixing function for arbitrary constrained dynamical
systems. This proves the independence of the vacuum functional under such
transformations. We derive a new form of the Ward identities, depending on the
parameters $\lambda_{a}$, and study the problem of gauge-dependence. We
present the form of transformation parameters which generates a change of the
gauge in the Hamiltonian path integral, evaluate it explicitly for connecting
two arbitrary $R_{\xi}$-like gauges in the Yang--Mills theory and establish,
after integration over momenta, a coincidence with the Lagrangian path
integral [arXiv:1405.0790[hep-th]], which justifies the unitarity of the
$S$-matrix in the Lagrangian approach.

\end{abstract}


\section{Introduction}

It is well known that modern quantization methods for gauge
theories in the Lagrangian and Hamiltonian formulations
\cite{books2,books1,books3,books4} are based mainly on the
principles of BRST symmetry \cite{BRST1,BRST2,BRST3} and
BRST-antiBRST symmetry \cite{aBRST1,aBRST2,aBRST3}, which are
characterized by the presence of one Grassmann-odd parameter $\mu$
and two Grassmann-odd parameters $(\mu,\bar{\mu})$, respectively.
The parameters of the $\mathrm{Sp}\left(  2\right)  $-covariant
generalized Hamiltonian \cite{BLT1h,BLT2h} and Lagrangian
\cite{BLT1,BLT2} quantization schemes (see also \cite{GH1,Hull})
form an $\mathrm{Sp}\left(  2\right)  $-doublet:
$(\mu,\bar{\mu})\equiv(\mu_{1},\mu_{2})=\mu_{a}$. These parameters
were initially considered as infinitesimal odd-valued objects and
may be regarded as constants and as field-dependent functionals,
used, respectively, to obtain the Ward identities and to establish
the gauge-independence of the corresponding vacuum functional in
the path integral approach.

In our recent work \cite{MRnew}, we have suggested an extension of
BRST-antiBRST transformations in Lagrangian formalism to
finite (both global and field-dependent) parameters in
Yang--Mills and general gauge theories, which in the latter case
has been recently developed in \cite{MRnew2, MRnew3}. The idea of
\textquotedblleft finiteness\textquotedblright\ is also based on
the inclusion into BRST-antiBRST transformations of a new term,
being quadratic in the transformation parameters $\lambda_{a}$.
This makes it possible to realize the complete BRST-antiBRST
invariance of the integrand in the vacuum functional. The
functionally-dependent parameters $\lambda_{a}=s_{a}\Lambda$,
induced by a Grassmann-even functional $\Lambda$, provide an
explicit correspondence (due to a so-called compensation
equation for the corresponding Jacobian) between the choices of
$\Lambda$ connecting the partition function of a theory in a
certain gauge (determined by a gauge Bosonic functional $F_{0}$)
with the theory in a different gauge, given by another gauge Boson
$F$. This becomes a key instrument to determine, in a
BRST-antiBRST approach, the Gribov horizon functional
\cite{Gribov} -- given by the Landau gauge in the
Gribov--Zwanziger theory \cite{Zwanziger} -- by using any other
gauge, including the $R_{\xi}$-gauges, which eliminate residual
gauge invariance in the deep IR region. Notice that the finite
BRST-antiBRST transformations are, in fact, constructed from
infinitesimal gauge transformations (instead of finite gauge group
transformations) of classical variables in the case of finite
values of gauge parameters. Therefore, finite BRST-antiBRST
transformations developed within perturbative theory may be used
to consistently\footnote{Namely, in a way that preserves the
gauge-independence of the physical $S$-matrix.} determine the
Gribov horizon functional in any differential gauge (due to
Singer's result \cite{Singer}), starting from the horizon
functional in a fixed gauge, which, in turn, should be obtained
non-perturbatively from finite gauge group transformations.

For the sake of completeness, let us remind that finite field-dependent BRST
transformations were introduced \cite{JM} in the Yang--Mills theory (with the
quantum action constructed by the Faddeev--Popov rules \cite{FP}), on the
basis of a functional equation for the parameter used to provide the path
integral with a change of variables that would allow one to relate the quantum
action in a certain gauge with the quantum action in a different gauge. This
equation and a similar equation \cite{RM} for the finite parameter of a
field-dependent BRST transformation in generalized Hamiltonian formalism
were solved in a series of particular cases for parameters;
however, a general solution was not presented.

The recent studies \cite{BLThf,Reshetnyak} have proposed the idea
of finite BRST--BFV transformations \cite{BLThf}\ in the
generalized Hamiltonian formalism \cite{BRST3,BFV,Henneaux1}, as
well as finite BRST \cite{Reshetnyak} and BRST--BV \cite{BLTfin}
transformations, using different path integral representations in
the Batalin--Vilkovisky (BV) formalism \cite{BV}. It has been
shown that, in order to relate partition functions given by
different gauges, it is sufficient to solve a compensation
equation for the corresponding finite field-dependent parameter,
first suggested in \cite{LL1} for Yang--Mills theories in the
Faddeev--Popov procedure \cite{FP}. This problem was raised
in \cite{llr1} to explore the issue of gauge-independence in gauge
theories with so-called soft breaking of BRST symmetry, which is
related to a consistent construction of the Gribov horizon
functional \cite{Zwanziger} by using different gauges \cite{LL2,
Reshetnyak2}.

Thus, the problem of setting up a construction of finite BRST-antiBRST
transformations for arbitrary dynamical systems with first-class constraints
and investigating its properties in generalized Hamiltonian formalism is open
even in Yang--Mills theories. This problem is related to establisphing a
correspondence of the quantum action in the BRST-antiBRST generalized
Hamiltonian quantization \cite{BLT1h,BLT2h} -- where gauge is introduced by a
Bosonic gauge-fixing function of phase-space variables $\Phi$ -- with the
quantum action of the same theory in a different gauge $\Phi+\Delta\Phi$ for a
finite value of $\Delta\Phi$, by using a change of variables in the path integral.

Based on these reasons, we intend to address the following issues in the case
of dynamical systems with first-class constraints in the generalized
Hamiltonian formalism:

\begin{enumerate}
\item introduction of \emph{finite BRST-antiBRST transformations}, being
polynomial in powers of a constant $\mathrm{Sp}\left(  2\right)  $-doublet of
Grassmann-odd parameters $\lambda_{a}$ and leaving the integrand in the
Hamiltonian path integral for vanishing external sources invariant to all
orders in $\lambda_{a}$;

\item definition of \emph{finite field-dependent BRST-antiBRST
transformations} as polynomials in the $\mathrm{Sp}\left(  2\right)  $-doublet
of Grassmann-odd functionals $\lambda_{a}(\Gamma)$ depending on the
entire set of symplectic coordinates of the total phase space; calculation of
the Jacobian related to this change of variables by using a special class of
transformations with $s_{a}$-potential parameters $\lambda_{a}(\Gamma)=\int
dt\ s_{a}\Lambda(\Gamma\left(  t\right)  )$, for a Grassmann-even function
$\Lambda(\Gamma\left(  t\right)  )$ and Grassmann-odd generators $s_{a}$ of
BRST-antiBRST transformations;

\item construction of a solution to the compensation equation for an unknown
function $\Lambda$ generating the $\mathrm{Sp}\left(  2\right)  $-doublet
$\lambda_{a}$ to establish a relation of the Hamiltonian action $S_{H,\Phi}$
in a certain gauge determined by a gauge Boson $\Phi$ with the Hamiltonian
action $S_{H,\Phi+\Delta\Phi}$ in a different gauge $\Phi+\Delta\Phi$;

\item explicit construction of the parameters $\lambda_{a}$ of finite
field-dependent BRST-antiBRST transformations generating a change of the gauge
in the Hamiltonian path integral within a class of linear $R_{\xi}$-like
gauges in the Hamiltonian formalism, which are realized in terms of Bosonic
gauge functions $\Phi_{\left(  \xi\right)  }$, with $\xi=0,1$ corresponding to
the Landau and Feynman (covariant) gauges, respectively.
\end{enumerate}

The work is organized as follows. In Section~\ref{gensetup}, we remind the
general setup of the BRST-antiBRST generalized Hamiltonian quantization of
dynamical systems with first-class constraints and list its basics
ingredients. In Section~\ref{fBRSTa}, we introduce the notion of finite
BRST-antiBRST transformations with constant and field-dependent
parameters in generalized Hamiltonian formalism. We obtain explicit Jacobians
corresponding to these changes of variables and show that, exactly as in the
case of field-dependent BRST-antiBRST transformations for the Yang--Mills
vacuum functional \cite{MRnew} in Lagrangian formalism, the corresponding
field-dependent transformations amount to a precise change of the gauge-fixing
functional. Here, we also study the group properties of finite field-dependent
BRST-antiBRST transformations. In Section~\ref{WIGD}, we derive the Ward
identities with the help of field-dependent BRST-antiBRST transformations and
study the gauge dependence of the generating functionals of Green's functions.
In Section~\ref{YMgauges}, we present the form of transformation parameters
that generates a change of the gauge and evaluate it for connecting two
arbitrary $R_{\xi}$-like gauges in Yang--Mills theories. In
Conclusion\label{concl}, we discuss the results and outline some open
problems. In Appendix~\ref{AppA}, we present a detailed calculation of the
Jacobians corresponding to the finite BRST-antiBRST Hamiltonian
transformations with constant and field-dependent parameters.

We use condensed notations similar to \cite{DeW}, namely, the spatial
coordinates of canonical field variables $\Gamma^{p}=\left(  P_{A}%
,Q^{A}\right) $ are absorbed into the indices $p$, $A$, whereas integration
over the spatial coordinates is included into summation over repeated indices.
The partial $\partial/\partial\Gamma^{p}$ and variational $\delta/\delta
\Gamma^{p}$ derivatives over $\Gamma^{p}$ are understood as acting from the
right. The variational derivative $\delta/\delta\Gamma^{p}\left(  t\right)  $
is taken along a phase-space trajectory $\Gamma^{p}\left(  t\right)  $,
whereas the partial derivative $\partial/\partial\Gamma^{p}$ of a field
variable $\Gamma^{p}$ is understood as the variational derivative with fixed
time, $\delta_{t}/\delta\Gamma^{p}$, as in \cite{books3}, applied to a
functional $\mathcal{F}\left(  \Gamma\left(  t\right)  \right)  $ local in
time, $\delta\mathcal{F}=\left(  \delta_{t}\mathcal{F}/\delta\Gamma
^{p}\right)  \delta\Gamma^{p}$, $\delta_{t}/\delta\Gamma^{p}\equiv
\partial/\partial\Gamma^{p}$. We refer to $t$-local functionals $\mathcal{F}%
\left(  \Gamma\right)  $ as \emph{functions}, whereas the corresponding
$F\left(  \Gamma\right)  =\int dt\ \mathcal{F}\left(  \Gamma\left(  t\right)
\right)  $ are called \emph{functionals}. The raising and lowering of
$\mathrm{Sp}\left(  2\right)  $ indices, $s^{a}=\varepsilon^{ab}s_{b}$,
$s_{a}=\varepsilon_{ab}s^{b}$, is carried out with the help of a constant
antisymmetric second-rank tensor $\varepsilon^{ab}$, $\varepsilon
^{ac}\varepsilon_{cb}=\delta_{b}^{a}$, subject to the normalization condition
$\varepsilon^{12}=1$. The Grassmann parity and ghost number of a quantity $A$,
assumed to be homogeneous with respect to these characteristics, are denoted
by $\varepsilon\left(  A\right)  $, $\mathrm{gh}(A)$, respectively. By
default, we understand BRST-antiBRST transformations in generalized
Hamiltonian formalism as \emph{infinitesimal} invariance transformations with
a doublet $\lambda_{a}$ of anticommuting parameters, whereas \emph{finite
BRST-antiBRST transformations }are understood as transformations of invariance
to all powers of the transformation parameters $\lambda_{a}$.

\section{Basics of BRST-antiBRST Generalized Hamiltonian Quantization}

\label{gensetup}
\renewcommand{\theequation}{\arabic{section}.\arabic{equation}} \setcounter{equation}{0}

We recall that the total phase space underlying the BRST-antiBRST generalized
Hamiltonian quantization is parameterized by the canonical
phase-space variables, $\Gamma^{p}$, $\varepsilon(\Gamma^{p})=\varepsilon_{p}%
$,
\begin{equation}
\Gamma^{p}=\left(  P_{A},Q^{A}\right)  =\left(  \eta,\Gamma_{\mathrm{gh}%
}\right)  \ , \label{tpsv}%
\end{equation}
where$\ \eta=\left(  p_{i},q^{i}\right)  $ are the classical momenta and
coordinates of a given dynamical system, described by a Hamiltonian
$H_{0}=H_{0}(\eta)$ and by a set of (generally, linearly dependent)
first-class constraints $T_{\alpha_{0}}=T_{\alpha_{0}}(\eta)$, $\varepsilon
(T_{\alpha_{0}})=\varepsilon_{\alpha_{0}}$, subject to involution relations in
terms of the Poisson superbracket at a fixed time instant $t$, $\{\Gamma
^{p},\Gamma^{q}\}=\omega^{pq}=\mathrm{const}$, with $\omega^{pq}$ being an
even supermatrix, $\omega^{pq}=-(-1)^{\varepsilon_{p}\varepsilon_{q}}%
\omega^{qp}$,%
\begin{equation}
\left\{  H_{0},T_{\alpha_{0}}\right\}  =T_{\gamma_{0}}V_{\alpha_{0}}%
^{\gamma_{0}},\quad\left\{  T_{\alpha_{0}},T_{\beta_{0}}\right\}
=T_{\gamma_{0}}U_{\alpha_{0}\beta_{0}}^{\gamma_{0}}\ ,\ \ \ \mathrm{for}%
\ \ \ \ U_{\alpha_{0}\beta_{0}}^{\gamma_{0}}=-(-1)^{\varepsilon_{\alpha_{0}%
}\varepsilon_{\beta_{0}}}U_{\beta_{0}\alpha_{0}}^{\gamma_{0}}\ .
\label{invrel}%
\end{equation}
The variables $\Gamma_{\mathrm{gh}}$ in (\ref{tpsv}) contain the entire set of
auxiliary variables that correspond to the towers \cite{BFV} of
ghost-antighost coordinates $C$ and Lagrangian multipliers $\pi$, as well as
their respective conjugate momenta $\mathcal{P}$ and $\lambda$, arranged
within the BRST-antiBRST generalized Hamiltonian quantization
\cite{BLT1h,BLT2h} into $\mathrm{Sp}(2)$-symmetric tensors for an $L$-th stage
of reducibility ($L=0$ corresponding to an irreducible theory),
\[
\Gamma_{\mathrm{gh}}=\left(  \mathcal{P}_{\alpha_{s}|a_{0}\ldots a_{s}%
}\ ,\ C^{\alpha_{s}|a_{0}...a_{s}},\ \lambda_{\alpha_{s}|a_{1}\ldots a_{s}%
},\ \pi^{\alpha_{s}|a_{1}...a_{s}}\ ,\ \ \ s=0,1,...,L\right)  \ ,
\]
with the corresponding distribution \cite{BLT2h} of the Grassmann parity and
ghost number.

The generating functional of Green's functions for a dynamical system in
question has the form%
\begin{equation}
Z_{\Phi}\left(  I\right)  =\int d\Gamma\exp\left\{  \frac{i}{\hbar}\int
dt\left[  \frac{1}{2}\Gamma^{p}(t)\omega_{pq}\dot{\Gamma}^{q}(t)-H_{\Phi
}(t)+I(t)\Gamma(t)\right]  \right\}  \label{ZPI}%
\end{equation}
and determines the partition function $Z_{\Phi}=Z_{\Phi}\left(  0\right)  $ at
the vanishing external sources $I_{p}(t)$ to ${\Gamma}^{p}$. In (\ref{ZPI}),
integration over time is taken over the range $t_{\mathrm{in}}\leq t\leq
t_{\mathrm{out}}$; the functions of time $\Gamma^{p}(t)\equiv\Gamma_{t}^{p}$
for $t_{\mathrm{in}}\leq t\leq t_{\mathrm{out}}$\ are trajectories,
$\dot{\Gamma}^{p}(t)\equiv d{\Gamma}^{p}(t)/dt$; the quantities $\omega
_{pq}=(-1)^{(\varepsilon_{p}+1)(\varepsilon_{q}+1)}\omega_{qp}$ compose an
even supermatrix inverse to that with the elements $\omega^{pq}$; the
unitarizing Hamiltonian $H_{\Phi}(t)=H_{\Phi}(\Gamma(t))$ is determined by
four $t$-local functions: $\mathcal{H}(t)$, an $\mathrm{Sp}(2)$-doublet of
odd-valued functions $\Omega^{a}(t)$, with $\mathrm{gh}(\Omega^{a})=-(-1)^{a}%
$, and an even-valued function $\Phi(t)$, with $\mathrm{gh}(\Phi)=0$, known as
the gauge-fixing Boson, which are given by the equations%
\begin{align}
&  H_{\Phi}(t)=\mathcal{H}(t)+\frac{1}{2}\varepsilon_{ab}\left\{  \left\{
\Phi(t),\Omega^{a}(t)\right\}  _{t},\Omega^{b}(t)\right\}  _{t}%
\ ,\ \ \mathrm{with}\ \ \left\{  A(t),B(t)\right\}  _{t}=\left.  \left\{
A(\Gamma),B(\Gamma)\right\}  \right\vert _{\Gamma=\Gamma(t)}%
\ ,\ \ \mathrm{for}\ \mathrm{any}\ A,B\ ,\label{Hphi}\\
&  \left\{  \Omega^{a},\Omega^{b}\right\}  =0\ ,\ \ \left\{  \mathcal{H}%
,\Omega^{b}\right\}  =0\ , \label{HOmega}%
\end{align}
with the boundary conditions%
\begin{equation}
\left.  \mathcal{H}\right\vert _{\Gamma_{\mathrm{gh}}=0}=H_{0}\left(
\eta\right)  \ ,\ \ \ \left.  \frac{\delta\Omega^{a}}{\delta C^{\alpha_{0}b}%
}\right\vert _{\Gamma_{\mathrm{gh}}=0}=\delta_{b}^{a}T_{\alpha_{0}}\left(
\eta\right)  \ . \label{bcond}%
\end{equation}
From equations (\ref{HOmega}) and the Jacobi identities for the Poisson
superbracket, it follows that%
\begin{equation}
\left\{  {H}_{\Phi},\Omega^{a}\right\}  =0\ . \label{HunitOm}%
\end{equation}
The integrand in (\ref{ZPI}) is invariant with respect to the infinitesimal
BRST-antiBRST transformations \cite{BLT1h}%
\begin{equation}
\Gamma^{p}\rightarrow\check{\Gamma}^{p}=\Gamma^{p}+\left(  s^{a}\Gamma
^{p}\right)  \mu_{a}\ ,\ \ \ \mathrm{with}\ \ \ \ s^{a}=\left\{
\bullet,\Omega^{a}\right\}  \ , \label{BABinf}%
\end{equation}
realized on phase-space trajectories $\Gamma^{p}(t)$ as%
\begin{equation}
\Gamma^{p}(t)\rightarrow\check{\Gamma}^{p}(t)=\Gamma^{p}(t)+\left\{
\Gamma^{p}(t),\Omega^{a}(t)\right\}  _{t}\mu_{a}=\Gamma^{p}(t)+\left(
s^{a}\Gamma^{p}\right)  \left(  t\right)  \mu_{a}\ , \label{BABinftr}%
\end{equation}
with an $\mathrm{Sp}(2)$-doublet $\mu_{a}$ of anticommuting constant
infinitesimal parameters, $\mu_{a}\mu_{b}+\mu_{a}\mu_{b}\equiv0$, for any
$a,b=1,2$. The generators $s^{a}$ of BRST-antiBRST transformations are
anticommuting, nilpotent and obey the Leibnitz rule when acting on the product
and the Poisson superbracket:%
\begin{equation}
s^{a}s^{b}+s^{b}s^{a}=0\ ,\ \ \ s^{a}s^{b}s^{c}=0\ ,\ \ \ s^{a}\left(
AB\right)  =\left(  s^{a}A\right)  B\left(  -1\right)  ^{\varepsilon_{B}%
}+A\left(  s^{a}B\right)  \ ,\ \ \ s^{a}\left\{  A,B\right\}  =\left\{
s^{a}A,B\right\}  \left(  -1\right)  ^{\varepsilon_{B}}+\left\{
A,s^{a}B\right\}  \ . \label{algPbr}%
\end{equation}
The BRST-antiBRST invariance of the integrand in (\ref{ZPI}) with $I_{p}(t)=0$
under the transformations (\ref{BABinftr}) allows one to obtain the Ward
identities for $Z_{\Phi}\left(  I\right)  $, namely,%
\begin{align}
&
\left\langle {}\right.  \int dt\ I_{p}(t)s^{a}\Gamma^{p}(t)\left.
{}\right\rangle _{\Phi,I}=0\,,
\label{WIham}\\
&  \mathrm{for}\,\,\,\langle\mathcal{O}\rangle_{\Phi,I}=Z_{\Phi}^{-1}(I)\int
d\Gamma\;\mathcal{O}\exp\left\{  \frac{i}{\hbar}\left[  S_{H,\Phi}%
(\Gamma)+\int dt\ I_{p}(t)\Gamma^{p}(t)\right]  \right\}  \ ,\nonumber\\
&  \mathrm{with}\,\,\,S_{H,\Phi}(\Gamma)=\int dt\left[  \frac{1}{2}\Gamma
^{p}(t)\omega_{pq}\dot{\Gamma}^{p}(t)-H_{\Phi}(t)\right]  \ , \label{expval}%
\end{align}
where the expectation value of a functional $\mathcal{O}(\Gamma)$ is
calculated with respect to a certain gauge $\Phi(\Gamma)$ in the presence of
external sources $I_{p}$\,. To obtain (\ref{WIham}), we subject (\ref{ZPI}) to a
change of variables $\Gamma\rightarrow\Gamma+\delta\Gamma$ with $\delta\Gamma$
given by (\ref{BABinftr}) and use the equations (\ref{HunitOm}) for $H(t)$. At
the same time, with allowance for the equivalence theorem \cite{equiv}, the
transformations (\ref{BABinftr}) allow one to establish the independence of
the $S$-matrix from the choice of a gauge. Indeed, if we change the gauge,
$\Phi\rightarrow\Phi+\Delta\Phi$, by an infinitesimal value $\Delta\Phi$ in
$Z_{\Phi}$ and make the change of variables (\ref{BABinftr}), choosing the
parameters $\mu_{a}$ as functionals of $\Gamma^{p}$ (i.e., not as functions of
time $t$ or of the variables $\Gamma^{p}$), namely,%
\begin{equation}
\mu_{a}=\frac{i}{2\hbar}\varepsilon_{ab}\int dt\left\{  \Delta\Phi
,\ \Omega^{b}\right\}  _{t}=\frac{i}{2\hbar}\int dt\ \left(  s_{a}\Delta
\Phi\right)  \left(  t\right)  \ , \label{inffdpar}%
\end{equation}
we arrive at $Z_{\Phi+\Delta\Phi}=Z_{\Phi}$, and therefore the $S$-matrix is gauge-independent.

\section{Finite BRST-antiBRST Transformations}

\label{fBRSTa} \renewcommand{\theequation}{\arabic{section}.\arabic{equation}} \setcounter{equation}{0}

In this section, we introduce (Subsection~\ref{deffin}) the notion of finite
BRST-antiBRST transformations and examine two classes of such transformation,
namely, those with constant and field-dependent parameters, each class
being realized in a $t$-local form and in a functional form. We calculate
(Subsection~\ref{jacobian}) the corresponding Jacobians, derive
(Subsection~\ref{compEq}) the compensation equation and present its solution.
Finally, we study (Subsection~\ref{grouprop}) some group properties of
field-dependent BRST-antiBRST transformations.

\subsection{Definitions}

\label{deffin}

Let us introduce finite transformations of the canonical variables $\Gamma
^{p}$ with a doublet $\lambda_{a}$ of anticommuting Grassmann parameters,
$\lambda_{a}\lambda_{b}+\lambda_{b}\lambda_{a}=0$,%
\begin{equation}
\Gamma^{p}\rightarrow\check{\Gamma}^{p}=\Gamma^{p}+\Delta\Gamma^{p}%
=\check{\Gamma}{}^{p}\left(  \Gamma|\lambda\right)
\ ,\ \ \ \mathrm{so\ \ that}\ \ \ \check{\Gamma}{}^{p}\left(  |0\right)
={\Gamma}{}^{p}\ . \label{defin}%
\end{equation}
In general, such transformations are quadratic in their parameters, due to
$\lambda_{a}\lambda_{b}\lambda_{c}\equiv0$,%
\begin{equation}
\check{\Gamma}{}^{p}\left(  \Gamma|\lambda\right)  =\check{\Gamma}{}%
^{p}\left(  \Gamma|0\right)  +\left[  \check{\Gamma}{}^{p}\left(
\Gamma|\lambda\right)  \frac{\overleftarrow{\partial}}{\partial\lambda_{a}%
}\right]  _{\lambda=0}\lambda_{a}+\frac{1}{2}\left[  \check{\Gamma}{}%
^{p}\left(  \Gamma|\lambda\right)  \frac{\overleftarrow{\partial}}%
{\partial\lambda_{a}}\frac{\overleftarrow{\partial}}{\partial\lambda_{b}%
}\right]  \lambda_{b}\lambda_{a}\ , \label{Z}%
\end{equation}
which implies%
\begin{equation}
\Delta\check{\Gamma}{}^{p}=Z^{pa}\lambda_{a}+\left(  1/2\right)  Z^{p}%
\lambda^{2}\,,\,\,\,\mathrm{where}\,\,\,\,\lambda^{2}\equiv\lambda_{a}%
\lambda^{a}\ , \label{Z2}%
\end{equation}
for certain functions $Z^{pa}=Z^{pa}\left(  \Gamma\right)  $, $Z^{p}%
=Z^{p}\left(  \Gamma\right)  $, corresponding to the first- and second-order
derivatives of $\check{\Gamma}{}^{p}\left(  \Gamma|\lambda\right)  $ with
respect to $\lambda_{a}$ in (\ref{Z}).

Let us consider an arbitrary function $\mathcal{F}(\Gamma)$ of phase-space
variables expandable as a series in powers of $\Gamma^{p}$. Because of the
nilpotency $\Delta\Gamma^{p_{1}}\cdots\Delta\Gamma^{p_{n}}\equiv0$, $n\geq3$,
the function $\mathcal{F}\left(  \Gamma\right)  $ under the transformations
(\ref{Z2}) can be expanded as%
\begin{equation}
\mathcal{F}\left(  \Gamma+\Delta\Gamma\right)  =\mathcal{F}\left(
\Gamma\right)  +\frac{\partial\mathcal{F}\left(  \Gamma\right)  }%
{\partial\Gamma^{p}}\Delta\Gamma^{p}+\frac{1}{2}\frac{\partial^{2}%
\mathcal{F}\left(  \Gamma\right)  }{\partial\Gamma^{p}\partial\Gamma^{q}%
}\Delta\Gamma^{q}\Delta\Gamma^{p}\ . \label{F}%
\end{equation}
Let the function $\mathcal{F}(\Gamma)$ be now invariant with respect to
infinitesimal BRST-antiBRST transformations (\ref{BABinf}),%
\begin{equation}
s^{a}\mathcal{F}(\Gamma)=0\ ,\ \ \ \mathrm{where}\ \ \ s^{a}\mathcal{F}%
(\Gamma)=\frac{\partial\mathcal{F}\left(  \Gamma\right)  }{\partial\Gamma^{p}%
}s^{a}\Gamma^{p}\ , \label{safcal}%
\end{equation}
and introduce \emph{finite BRST-antiBRST transformations} in generalized
Hamiltonian formalism as invariance transformations of the function
$\mathcal{F}(\Gamma)$ under finite transformations of the variables
$\Gamma^{p}$, such that%
\begin{equation}
\mathcal{F}\left(  \Gamma+\Delta\Gamma\right)  =\mathcal{F}\left(
\Gamma\right)  \ ,\ \ \ \left.  \Delta\Gamma^{p}\frac{\overleftarrow{\partial
}}{\partial\lambda_{a}}\right\vert _{\lambda=0}=s^{a}\Gamma^{p}%
\ \ \ \mathrm{and}\mathtt{\ \ \ }\Delta\Gamma^{p}\frac{\overleftarrow
{\partial}}{\partial\lambda_{a}}\frac{\overleftarrow{\partial}}{\partial
\lambda_{b}}=-\frac{1}{2}\varepsilon^{ab}s^{2}\Gamma^{p}%
\ ,\ \ \ \mathrm{where}\,\,\,\,s^{2}\equiv s_{a}s^{a}\ .
\label{finBRSTantiBRST}%
\end{equation}
Namely, for the transformed variables $\check{\Gamma}^{p}=\Gamma^{p}%
+\Delta\Gamma^{p}$ we have\footnote{As shown in \cite{MRnew}, the
validity of the algebra of BRST-antiBRST transformations for its
generators
$\overleftarrow{s}^a\overleftarrow{s}^b+\overleftarrow{s}^b
\overleftarrow{s}^a=0$, realized in an appropriate space of
variables in Lagrangian \cite{BLT1} and generalized Hamiltonian
formalism \cite{BLT2h} allows one to restore the finite group form
$\check{\Gamma}-\Gamma=\Gamma\left(\overleftarrow{s}^a\lambda_a +
(1/4)\overleftarrow{s}^2\lambda^2\right)$, or, identically,
$\check{\Gamma}=\Gamma\left(1+\overleftarrow{s}^a\lambda_a +
(1/4)\overleftarrow{s}^2\lambda^2\right)=\Gamma\exp\left(\overleftarrow{s}^a\lambda_a\right)$.
Equivalently, the realization of the generators in terms of
odd-valued anticommuting vector fields,
$\overleftarrow{s}^a\left(\Gamma\right)=\frac{\overleftarrow{\delta}}{\delta
\Gamma^p} (\Gamma^p\overleftarrow{s}^a)$, due to the Frobenius
theorem, leads to the same form of finite BRST-antiBRST
transformations.
}%
\begin{equation}
\check{\Gamma}^{p}=\Gamma^{p}\left(  1+\overleftarrow{s}^{a}\lambda_{a}%
+\frac{1}{4}\overleftarrow{s}^{2}\lambda^{2}\right)
\ ,\ \ \mathrm{or,\ equivalently,}\ \ \Delta\Gamma^{p}=\left(  s^{a}\Gamma
^{p}\right)  \lambda_{a}+\frac{1}{4}\left(  s^{2}\Gamma^{p}\right)
\lambda^{2}\,,\,\,\,\mathrm{where}\,\,\,\,\overleftarrow{s}^{2}\equiv
\overleftarrow{s}^{a}\overleftarrow{s}_{a}\ , \label{deffin1}%
\end{equation}
which is realized on phase-space trajectories $\Gamma^{p}(t)$ as follows:%
\begin{equation}
\check{\Gamma}^{p}\left(  t\right)  =\Gamma^{p}\left(  t\right)  \left(
1+\overleftarrow{s}^{a}\lambda_{a}+\frac{1}{4}\overleftarrow{s}^{2}\lambda
^{2}\right)  \ ,\ \ \mathrm{or,\ eqiuvalently,}\ \ \Delta\Gamma^{p}\left(
t\right)  =\left(  s^{a}\Gamma^{p}\right)  \left(  t\right)  \lambda_{a}%
+\frac{1}{4}\left(  s^{2}\Gamma^{p}\right)  \left(  t\right)  \lambda^{2}\,.
\label{deffin_traj}%
\end{equation}

Let us now consider an arbitrary functional of the phase-space variables,
$F(\Gamma)$, expandable as a series in powers of $\Gamma^{p}$. Under the
transformations (\ref{BABinftr}), the functional $F\left(  \Gamma\right)  $
can be presented as%
\begin{equation}
F\left(  \Gamma+\Delta\Gamma\right)  =F\left(  \Gamma\right)  +\int
dt\ \frac{\delta F\left(  \Gamma\right)  }{\delta\Gamma^{p}\left(  t\right)
}\Delta\Gamma^{p}\left(  t\right)  +\frac{1}{2}\int dt^{\prime}\ dt^{\prime
\prime}\frac{\delta^{2}F\left(  \Gamma\right)  }{\delta\Gamma^{p}\left(
t^{\prime}\right)  \delta\Gamma^{q}\left(  t^{\prime\prime}\right)  }%
\Delta\Gamma^{q}\left(  t^{\prime\prime}\right)  \Delta\Gamma^{p}\left(
t^{\prime}\right)  \ . \label{def_func}%
\end{equation}
By analogy with the definition (\ref{safcal}) of BRST-antiBRST transformations
of functions, we let the functional $F(\Gamma)$ be invariant with respect to
infinitesimal BRST-antiBRST transformations for trajectories (\ref{BABinftr}),%
\begin{equation}
s^{a}F(\Gamma)=0\ ,\ \ \ \mathrm{where}\ \ \ s^{a}F(\Gamma)=\int
dt\ \frac{\delta F\left(  \Gamma\right)  }{\delta\Gamma^{p}\left(  t\right)
}\left(  s^{a}\Gamma^{p}\right)  \left(  t\right)  \ , \label{safcal2}%
\end{equation}
and introduce the \emph{finite BRST-antiBRST transformations of functionals}
as invariance transformations of a functional $F(\Gamma)$ under finite
transformations of trajectories $\Gamma^{p}\left(  t\right)  \rightarrow
\check{\Gamma}^{p}\left(  t\right)  $, such that%
\begin{equation}
F(\check{\Gamma})=F\left(  \Gamma\right)  \ ,\ \ \ \check{\Gamma}^{p}\left(
t\right)  =\Gamma^{p}\left(  t\right)  \left(  1+\overleftarrow{s}^{a}%
\lambda_{a}+\frac{1}{4}\overleftarrow{s}^{2}\lambda^{2}\right)  \ .
\label{finBRSTantiBRST2}%
\end{equation}
The definitions of finite BRST-antiBRST transformations realized on functions
(\ref{deffin_traj}) and functionals (\ref{finBRSTantiBRST2}) are consistent.
Indeed, for an arbitrary function $\mathcal{F}(t)=\mathcal{F}(\Gamma(t))$ with
the corresponding functional ${F}(\Gamma)=\int dt\ \mathcal{F}(t)$, we have%
\begin{align}
s^{a}F(\Gamma)  &  =\int dt\ \frac{\delta F\left(  \Gamma\right)  }%
{\delta\Gamma^{p}\left(  t\right)  }\left(  s^{a}\Gamma^{p}\right)  \left(
t\right)  =\int dt\ \frac{\partial\mathcal{F}\left(  t\right)  }%
{\partial\Gamma^{p}\left(  t\right)  }\left(  s^{a}\Gamma^{p}\right)  \left(
t\right)  =\int dt\ s^{a}\mathcal{F}(t)\label{safcal3}\\
&  \Longrightarrow\Delta F(\Gamma)=\int dt\ \Delta\mathcal{F}\left(
\Gamma(t)\right)  \ ,\ \ \ \mathrm{with\ \ \ }\ \Delta F(\Gamma)=F(\check
{\Gamma})-F(\Gamma),\quad\Delta\mathcal{F}(\Gamma(t))=\mathcal{F}%
(\check{\Gamma}(t))-\mathcal{F}(\Gamma(t))\ . \label{safcal4}%
\end{align}
Formula (\ref{safcal3}) describes the rule according to which the
generators\footnote{To be more exact, one could use two different symbols for
the generators $s^{a}$ as they act on functions and functionals in (\ref{safcal}),
(\ref{safcal2}); however, in order to simplify the notation for virtually the
same operation, in view of (\ref{safcal3}), we use the symbol $s^{a}$.}
$s^{a}$ of BRST-antiBRST transformations act on functionals via functions
given in the phase space of $\Gamma^{p}$.

The consistency of definitions (\ref{deffin1}), (\ref{deffin_traj}),
(\ref{finBRSTantiBRST2}) is readily established by considering the respective
equations $\Delta\mathcal{F}=0$, $\Delta\mathcal{F}(t)=0$, $\Delta{F}=0$. For
the first equation, we have%
\begin{equation}
\frac{\partial\mathcal{F}\left(  \Gamma\right)  }{\partial\Gamma^{p}}\left[
(s^{a}\Gamma^{p})\lambda_{a}+\frac{1}{4}\left(  s^{2}\Gamma^{p}\right)
\lambda^{2}\right]  +\frac{1}{2}\frac{\partial^{2}\mathcal{F}\left(
\Gamma\right)  }{\partial\Gamma^{p}\partial\Gamma^{q}}\left[  (s^{a}\Gamma
^{q})\lambda_{a}+\frac{1}{4}\left(  s^{2}\Gamma^{q}\lambda^{2}\right)
\right]  \left[  (s^{b}\Gamma^{p})\lambda_{b}+\frac{1}{4}\left(  s^{2}%
\Gamma^{p}\right)  \lambda^{2}\right]  =0\ . \label{compbab}%
\end{equation}
Taking into account the fact that $\lambda_{a}\lambda^{2}=\lambda^{4}\equiv0$,
the invariance relations $s^{a}F(\Gamma)=\left(  \partial\mathcal{F}%
/\partial\Gamma^{p}\right)  s^{a}\Gamma^{p}=0$, and their differential
consequence (after applying $s^{b}$ and multiplying by $\lambda_{b}\lambda
_{a}$)%
\begin{equation}
\frac{\partial^{2}\mathcal{F}\left(  \Gamma\right)  }{\partial\Gamma
^{p}\partial\Gamma^{q}}(s^{b}\Gamma^{q})\lambda_{b}(s^{a}\Gamma^{p}%
)\lambda_{a}=-\frac{1}{2}\frac{\partial\mathcal{F}\left(  \Gamma\right)
}{\partial\Gamma^{p}}\left(  s^{2}\Gamma^{p}\right)  \lambda^{2},
\label{diffrel}%
\end{equation}
in view of the definition (\ref{BABinf}) and properties (\ref{algPbr}), we
find that the above equation (\ref{compbab}) is satisfied identically:%
\begin{equation}
\frac{\partial\mathcal{F}\left(  \Gamma\right)  }{\partial\Gamma^{p}}%
(s^{a}\Gamma^{p})\lambda_{a}+\frac{1}{4}\frac{\partial\mathcal{F}\left(
\Gamma\right)  }{\partial\Gamma^{p}}\left(  s^{2}\Gamma^{p}\right)
\lambda^{2}+\frac{1}{2}\frac{\partial^{2}\mathcal{F}\left(  \Gamma\right)
}{\partial\Gamma^{p}\partial\Gamma^{q}}(s^{b}\Gamma^{q})\lambda_{b}%
(s^{a}\Gamma^{p})\lambda_{a}\overset{(\ref{diffrel})}{\equiv}0\ .
\label{identf}%
\end{equation}
In a similar way, one can readily establish the consistency of definitions
(\ref{deffin_traj}) and (\ref{finBRSTantiBRST2}).

We can see that the finite variation $\Delta\Gamma^{p}$ includes the
generators of BRST-antiBRST transformations $\left(  s^{1},s^{2}\right)  $, as
well as their commutator $s^{2}=\varepsilon_{ab}s^{b}s^{a}=s^{1}s^{2}%
-s^{2}s^{1}$. According to (\ref{finBRSTantiBRST}), (\ref{def_func}) and
$\lambda_{a}\lambda^{2}=\lambda^{4}\equiv0$, the variations $\Delta
\mathcal{F}\left(  \Gamma\right)  $, $\Delta F\left(  \Gamma\right)  $ of an
arbitrary function $\mathcal{F}\left(  \Gamma\right)  $ and of an arbitrary
functional $F\left(  \Gamma\right)  $ under the corresponding finite
BRST-antiBRST transformations (\ref{deffin1}), (\ref{finBRSTantiBRST2}) are
given by%
\begin{equation}
\Delta\mathcal{F}=\left(  s^{a}\mathcal{F}\right)  \lambda_{a}+\frac{1}%
{4}\left(  s^{2}\mathcal{F}\right)  \lambda^{2}\quad\mathrm{and}\quad\Delta
F=\left(  s^{a}F\right)  \lambda_{a}+\frac{1}{4}\left(  s^{2}F\right)
\lambda^{2}\ . \label{DelFfrule}%
\end{equation}
In particular, the functions $\Omega^{a}\ $and $\mathcal{H}$ obey finite
BRST-antiBRST invariance:%
\begin{equation}
\Delta\Omega^{a}=\{\Omega^{a},\Omega^{b}\}\lambda_{b}+\frac{1}{4}%
\varepsilon_{bc}\left\{  \Omega^{a},\{\Omega^{b},\Omega^{c}\}\right\}
\lambda^{2}=0\ ,\ \ \ \Delta\mathcal{H}=\{\mathcal{H},\Omega^{a}\}\lambda
_{a}+\frac{1}{4}\varepsilon_{ab}\left\{  \mathcal{H},\{\Omega^{a},\Omega
^{b}\}\right\}  \lambda^{2}=0\ , \label{consomh}%
\end{equation}
due to the generating equations (\ref{HOmega}), with the corresponding
property for the Hamiltonian action $S_{H}(\Gamma)$ in (\ref{expval})
\begin{equation}
\Delta S_{H}(\Gamma)=S_{H}(\check{\Gamma})-S_{H}(\Gamma)=\int dt\left[
\frac{1}{2}\left(  \check{\Gamma}^{p}\omega_{pq}\frac{d\check{\Gamma}^{p}}%
{dt}\right)  (t)-H_{\Phi}\left(  \check{\Gamma}\right)  (t)\right]
-S_{H}(\Gamma)= \int dt\, \frac{d\mathcal{F}(t)}{dt} \ , \label{varSH}%
\end{equation}
where we have used the finite BRST-antiBRST invariance (\ref{consomh}) of the
unitarizing Hamiltonian $H_{\Phi}$ and the following transformations of the
term $\left(  {1}/{2}\right)  \int dt\, ( {\Gamma}^{p}\omega_{pq}\dot{\Gamma
}^{q}) $ with respect to the BRST-antiBRST transformations (\ref{deffin_traj})
of trajectories $\Gamma^{p}(t)$ leading to the appearance of ${d\mathcal{F}%
(t)}/dt $:%
\begin{equation}
\frac{1}{2}\int dt\left(  \check{\Gamma}^{p}\omega_{pq}\frac{d\check{\Gamma
}^{p}}{dt}\right)  (t)=\frac{1}{2}\left.  \left[  \left(  \Gamma^{p}%
\partial_{p}\Omega^{a}-2\Omega^{a}\right)  \lambda_{a}+\frac{1}{4}\Gamma
^{p}s_{a}(\partial_{p}\Omega^{a})\lambda^{2}\right]  \right\vert
_{t_{\mathrm{in}}}^{t_{\mathrm{out}}}+\frac{1}{2}\int dt\ \left(  {\Gamma}%
^{p}\omega_{pq}\dot{\Gamma}^{q}\right)  (t)\ , \label{pottrans}%
\end{equation}
which reflects the equality of the action in terms of the new phase-space
coordinates $\check{\Gamma}$ to the action in terms of the old coordinates $\Gamma$
up to a total derivative. The parameters $\lambda_{a}$ in (\ref{deffin1}),
(\ref{deffin_traj}) and (\ref{finBRSTantiBRST2}) may be constant, $\lambda
_{a}=\mathrm{const}$, as well as field-dependent, $\lambda_{a}=\lambda
_{a}(\Gamma)$, thus determining \emph{global }and\emph{ field-dependent finite
BRST-antiBRST transformations}. At the same time, we emphasize that the
parameters $\lambda_{a}(\Gamma)$ are not regarded as functions of time $t$,
and therefore of phase-space variables $\Gamma^{p}$, namely,%
\begin{equation}
\frac{d\lambda_{a}(\Gamma)}{dt}=\frac{\partial\lambda_{a}(\Gamma)}%
{\partial\Gamma^{p}}=0\ ;\ \ \mathrm{however,}\ \ \frac{\delta\lambda
_{a}(\Gamma)}{\delta\Gamma^{p}}\not \equiv 0\ . \label{condtglam}%
\end{equation}
Relations (\ref{deffin_traj}) and (\ref{DelFfrule}) allow one to calculate the
Jacobians of finite BRST-antiBRST transformations, as well as to investigate
the group properties of finite BRST-antiBRST transformations, presented in
respective Subsections~\ref{jacobian}, \ref{grouprop}. Thus, the functional
measure $d\Gamma$ in (\ref{ZPI}) turns out to be invariant with respect to the
change of trajectories, $\Gamma^{p}(t)\rightarrow\check{\Gamma}{}^{p}(t)$,
related to finite BRST-antiBRST transformations (\ref{deffin1}) with constant
parameters $\lambda_{a}$. This is nothing else than Liouville's theorem for
the transformations (\ref{deffin1}), being canonical, due to the identity%
\begin{equation}
\check{P}_{A}d\check{Q}^{A}-\check{H}_{\Phi}\left(  \check{P},\check
{Q}\right)  dt=P_{A}dQ^{A}-H_{\Phi}\left(  P,Q\right)  dt+d\mathcal{F}\ ,
\label{defcan}%
\end{equation}
which takes place for the contact $1$-form, as one makes the substitution
$\Gamma\rightarrow\check{\Gamma}$, setting $\check{H}_{\Phi}(\check{\Gamma
})=H_{\Phi}\left(  \check{\Gamma}\right)  $ and taking account of
(\ref{varSH}). The invariance of the measure, $d\check{\Gamma}=d\Gamma$, along
with the invariance (\ref{varSH}) of the action $S_{H}(\Gamma)$, justifies the
term \textquotedblleft finite BRST-antiBRST transformations\textquotedblright%
\ as applied to the invariance transformations (\ref{finBRSTantiBRST2}) of the
integrand for $Z_{\Phi}$.

\subsection{Jacobians}

\label{jacobian}

Let us examine the change of the integration measure $d\Gamma\rightarrow
d\check{\Gamma}$ in (\ref{ZPI}) under the finite transformations of
phase-space trajectories, $\Gamma_{t}^{p}\rightarrow\check{\Gamma}_{t}%
^{p}=\Gamma_{t}^{p}+\Delta\Gamma_{t}^{p}$, with $\Delta\Gamma_{t}^{p}%
\equiv\Delta\Gamma^{p}\left(  t\right)  $ given by (\ref{deffin_traj}),%
\begin{equation}
d\check{\Gamma}=d\Gamma\ \mathrm{Sdet}\left(  \frac{\delta\check{\Gamma}%
}{\delta\Gamma}\right)  ,\,\,\,\mathrm{Sdet}\left(  \frac{\delta\check{\Gamma
}}{\delta\Gamma}\right)  =\mathrm{Sdet}\left(  \mathbb{I}+M\right)
=\exp\left[  \mathrm{Str}\ln\left(  \mathbb{I}+M\right)  \right]  \equiv
\exp\left(  \Im\right)  \ , \label{measure}%
\end{equation}
where the Jacobian $\exp\left(  \Im\right)  $ has the form%
\begin{align}
\Im &  =\mathrm{Str}\ln\left(  \mathbb{I}+M\right)  =-\sum_{n=1}^{\infty}%
\frac{\left(  -1\right)  ^{n}}{n}\,\,\mathrm{Str}\left(  M^{n}\right)
,\,\,\,\mathrm{\ Str}\left(  M^{n}\right)  =\left(  -1\right)  ^{\varepsilon
_{p}}\int dt\ \left(  M^{n}\right)  _{p}^{p}\left(  t,t\right)  \,,\nonumber\\
\mathbb{I}  &  \mathbb{=\delta}_{q}^{p}\delta\left(  t^{\prime}-t^{\prime
\prime}\right)  \ ,\ \ \ \left(  M\right)  _{q}^{p}\left(  t^{\prime
},t^{\prime\prime}\right)  =\frac{\delta\Delta\Gamma^{p}\left(  t^{\prime
}\right)  }{\delta\Gamma^{q}\left(  t^{\prime\prime}\right)  }\ ,\ \ \ \left(
AB\right)  _{q}^{p}\left(  t^{\prime},t^{\prime\prime}\right)  =\int
dt\ \left(  A\right)  _{r}^{p}\left(  t^{\prime},t\right)  B_{q}^{r}\left(
t,t^{\prime\prime}\right)  \ . \label{superJ}%
\end{align}
In the case of finite transformations corresponding to $\lambda_{a}%
=\mathrm{const}$, the integration measure remains invariant (for details, see
(\ref{const}) in Appendix~\ref{AppA})
\begin{equation}
\Im\left(  \Gamma\right)  =0\Longrightarrow\left[  \mathrm{Sdet}\left(
\frac{\delta\check{\Gamma}}{\delta\Gamma}\right)  =1\,,\ \ \ d\check{\Gamma
}=d\Gamma\right]  \ . \label{constJ}%
\end{equation}
As we turn to finite field-dependent transformations, $\lambda_{a}=\lambda
_{a}\left(  \Gamma\right)  $, let us examine the particular case of
functionally-dependent parameters\footnote{The parameters $\lambda_{a}$ are
functionally-dependent, since $s^{1}\lambda_{1}+s^{2}\lambda_{2}=-\int
dt\ s^{2}\Lambda$.}%
\begin{equation}
\lambda_{a}\left(  \Gamma\right)  =\int dt\ \left(  s_{a}\Lambda\right)
\left(  t\right)  =\varepsilon_{ab}\int dt\ \left\{  \Lambda\left(  t\right)
,\Omega^{b}\left(  t\right)  \right\}  _{t}\ , \label{funcdepla}%
\end{equation}
with a certain even-valued potential function $\Lambda\left(  t\right)
=\Lambda\left(  \Gamma\left(  t\right)  \right)  $, which is inspired by
field-dependent BRST-antiBRST transformations with the parameters
(\ref{inffdpar}). In this case, the integration measure takes the form (for
details see (\ref{potent}) in Appendix~\ref{AppA})%
\begin{align}
&  \Im\left(  \Gamma\right)  =-2\mathrm{\ln}\left[  1+f\left(  \Gamma\right)
\right]  \ ,\,\,\,\mathrm{\ }\,f\left(  \Gamma\right)  =-\frac{1}{2}\int
dt\ \left(  s^{2}\Lambda\right)  _{t}\ ,\ \ \ \left(  s^{2}\Lambda\right)
_{t}=\varepsilon_{ab}\left\{  \left\{  \Lambda,\Omega^{a}\right\}  _{t}%
,\Omega^{b}\right\}  _{t}\ ,\label{superJaux}\\
&  d\check{\Gamma}=d\Gamma\exp\left[  \frac{i}{\hbar}\left(  -i\hbar
\Im\right)  \right]  =d\Gamma\exp\left\{  \frac{i}{\hbar}\left[
i\hbar\,\mathrm{\ln}\left(  1-\frac{1}{2}\varepsilon_{ab}\int dt\left\{
\left\{  \Lambda,\Omega^{a}\right\}  _{t},\Omega^{b}\right\}  _{t}\right)
^{2}\right]  \right\}  \ . \label{superJ1}%
\end{align}

\subsection{Solution of the Compensation Equation}

\label{compEq}Let us apply the Jacobian (\ref{superJ1}) to cancel a change of
the gauge Boson $\Phi(\Gamma)$ in (\ref{expval}):%
\begin{equation}
\Phi\rightarrow\Phi+\Delta\Phi\ . \label{changePhi}%
\end{equation}
To this end, we subject $Z_{\Phi+\Delta\Phi}$ to a change of variables
$\Gamma^{p}(t)\rightarrow\check{\Gamma}{}^{p}(t)$, given by (\ref{deffin_traj}%
) and parameterized by $\lambda_{a}\left(  \Gamma\right)  $ in accordance with
(\ref{funcdepla}). In terms of the new variables, we have%
\begin{align}
Z_{\Phi+\Delta\Phi}  &  =\int d\check{\Gamma}\exp\left[  \frac{i}{\hbar
}S_{H,\Phi+\Delta\Phi}(\check{\Gamma})\right]  =\int d{\Gamma}\exp\left[
{\Im\left(  \Gamma\right)  }\right]  \exp\left[  \frac{i}{\hbar}%
S_{H,\Phi+\Delta\Phi}({\Gamma})\right] \nonumber\\
&  =\int d{\Gamma}\exp\left[  {\Im\left(  \Gamma\right)  }\right]
\exp\left\{  \frac{i}{\hbar}\left[  S_{H,\Phi}({\Gamma})-\frac{1}%
{2}\varepsilon_{ab}\int dt\left\{  \left\{  \Delta\Phi(t),\Omega
^{a}(t)\right\}  _{t},\Omega^{b}(t)\right\}  _{t}\right]  \right\}  \ ,
\label{explcompeq}%
\end{align}
using the transformation property (\ref{varSH}) for $S_{H,\Phi+\Delta\Phi}$.
If we now require the fulfillment of the relation%
\begin{equation}
\exp\left[  {\Im\left(  \Gamma\right)  }\right]  =\exp{\left[  \frac{i}%
{2\hbar}\varepsilon_{ab}\int dt\left\{  \left\{  \Delta\Phi(t),\Omega
^{a}(t)\right\}  _{t},\Omega^{b}(t)\right\}  _{t}\right]  \ }, \label{eqexpl}%
\end{equation}
which we will call the \textquotedblleft compensation
equation\textquotedblright, then%
\begin{equation}
Z_{\Phi+\Delta\Phi}=Z_{\Phi}\ . \label{zphizphi1}%
\end{equation}
Using the relation (\ref{superJ1}) and the compensation equation (\ref{eqexpl})%
\begin{equation}
\frac{1}{2}\int dt\ \varepsilon_{ab}\left\{  \left\{  \Lambda,\Omega
^{a}\right\}  _{t},\Omega^{b}\right\}  _{t}=1-\exp{\left[  {\frac{1}%
{\,4i\hbar}}\varepsilon_{ab}\int dt\left\{  \left\{  \Delta\Phi(t),\Omega
^{a}(t)\right\}  _{t},\Omega^{b}(t)\right\}  _{t}\right]  }\ ,
\label{eqexpllam}%
\end{equation}
we can see that this is a functional equation for an unknown Bosonic function
$\Lambda(\Gamma)$, which determines $\lambda_{a}\left(  \Gamma\right)  $ in
accordance with $\lambda_{a}\left(  \Gamma\right)  =\int dt\ s_{a}%
\Lambda(\Gamma)$.

Introducing an auxiliary functional $y(\Gamma)$,%
\begin{equation}
y(\Gamma)\equiv\frac{1}{4i\hbar}\varepsilon_{ab}\int dt\left\{  \left\{
\Delta\Phi(t),\Omega^{a}(t)\right\}  _{t},\Omega^{b}(t)\right\}  _{t}=\frac
{1}{4i\hbar}\Delta\widehat{\Phi}\overleftarrow{s}^{2}\ ,\ \ \ \mathrm{where}%
\ \ \Delta\widehat{\Phi}\equiv\int dt\ \Delta{\Phi}(t)\ , \label{y}%
\end{equation}
which is BRST-antiBRST exact, $y(\Gamma)\overleftarrow{s}^{a}=0$, and making
use of $\overleftarrow{s}^{2}=\overleftarrow{s}^{a}\overleftarrow{s}_{a}$,
where $\left(  F\overleftarrow{s}^{a}\right)  \left(  \Gamma\right)  $ is
identical with $s^{a}F\left(  \Gamma\right)  $ in (\ref{safcal2}), we present
(\ref{eqexpllam}) in the form%
\begin{equation}
\frac{1}{2}\int dt\ \Lambda\overleftarrow{s}^{2}=1-\exp\left(  y\right)
=\frac{1}{4i\hbar}\left[  g(y)\Delta\widehat{\Phi}\right]  \overleftarrow
{s}^{2}\ , \label{compeq2}%
\end{equation}
where $g(y)=\left[  1-\exp(y)\right]  /y$ is a BRST-antiBRST exact functional.
This provides an explicit solution of (\ref{compeq2}), with accuracy up to
BRST-antiBRST exact terms:%
\begin{equation}
\Lambda(\Gamma|\Delta{\Phi})=\frac{1}{2i\hbar}g(y)\Delta{\Phi}\ .
\label{solcompeq2}%
\end{equation}
Hence, the field-dependent parameters $\lambda_{a}\left(  \Gamma\right)  $ are
implied by (\ref{funcdepla}) and (\ref{solcompeq2}),
\begin{equation}
\lambda_{a}(\Gamma|\Delta{\Phi})=\frac{1}{2i\hbar}g(y)\int dt\ \left(
s_{a}\Delta{\Phi}\right)  \left(  t\right)  =\frac{1}{2i\hbar}\varepsilon
_{ab}g(y)\int dt\ \left\{  \Delta{\Phi}(t),\Omega^{b}\left(  t\right)
\right\}  _{t}\ , \label{funcdeplafin}%
\end{equation}
whereas the approximation linear in $\Delta\Phi$ follows from $g\left(
0\right)  =-1$,%
\begin{equation}
\Lambda(\Gamma)=\frac{i}{2\hbar}\Delta{\Phi}+o\left(  \Delta{\Phi}\right)
\Longrightarrow\lambda_{a}(\Gamma)=\frac{i}{2\hbar}\varepsilon_{ab}\int
dt\ \left\{  \Delta{\Phi}(t),\Omega^{b}\left(  t\right)  \right\}
_{t}+o\left(  \Delta{\Phi}\right)  \ , \label{solcompeq3}%
\end{equation}
and is identical with the parameters (\ref{inffdpar}) of infinitesimal
field-dependent BRST-antiBRST transformations.

\subsection{Group Properties}

\label{grouprop}

The above relations (\ref{DelFfrule})%
\[
\Delta\mathcal{F}=\left(  s^{a}\mathcal{F}\right)  \lambda_{a}+\frac{1}%
{4}\left(  s^{2}\mathcal{F}\right)  \lambda^{2}\ ,\quad\Delta F=\left(
s^{a}F\right)  \lambda_{a}+\frac{1}{4}\left(  s^{2}F\right)  \lambda^{2}\ ,
\]
describing the finite variations of functions, $\mathcal{F=F}\left(
\Gamma\left(  t\right)  \right)  $, and functionals, $F=F\left(
\Gamma\right)  $, induced by finite BRST-antiBRST transformations, allow one
to study the group properties of these transformations, with the provision
that the transformations do not form neither a Lie superalgebra nor a
vector superspace, due to the quadratic dependence on the parameters
$\lambda_{a}$.

Let us study the composition of finite variations $\Delta_{\left(  1\right)
}\Delta_{\left(  2\right)  }$ acting on an object $A\left(  \Gamma\right)  $
being an arbitrary function or a functional. Using the Leibnitz-like
properties of the generators of BRST-antiBRST transformations, $s^{a}$ and
$s^{2}$, acting on the product of any functions (functionals) $A$, $B$ with
definite Grassmann parities,%
\begin{align}
&  s^{a}\left(  AB\right)  =\left(  s^{a}A\right)  B\left(  -1\right)
^{\varepsilon_{B}}+A\left(  s^{a}B\right)  \,,\,\,\,\,s_{a}\left(  AB\right)
=\left(  s_{a}A\right)  B\left(  -1\right)  ^{\varepsilon_{B}}+A\left(
s_{a}B\right)  \,,\nonumber\\
&  s^{2}\left(  AB\right)  =\left(  s^{2}A\right)  B-2\left(  s_{a}A\right)
\left(  s^{a}B\right)  \left(  -1\right)  ^{\varepsilon_{B}}+A\left(
s^{2}B\right)  \,,\, \label{s2(AB)}%
\end{align}
and the identities%
\begin{equation}
s^{a}s^{b}=\left(  1/2\right)  \varepsilon^{ab}s^{2}\,\,\,\,\mathrm{and}%
\,\,\,\,s_{a}s^{b}=-s^{b}s_{a}=\left(  1/2\right)  \delta_{a}^{b}%
s^{2}\,\,\,\,\mathrm{and}\,\,\,\,s^{a}s^{b}s^{c}\equiv0\ , \label{sasb}%
\end{equation}
with the notation $UV\equiv U_{a}V^{a}=-U^{a}V_{a}$ for pairing up any
$\mathrm{Sp}(2)$-vectors $U^{a}$, $V^{a}$, we obtain%
\begin{align}
s^{a}\left(  \Delta A\right)   &  =s^{a}\left[  \left(  s^{b}A\right)
\lambda_{b}+\frac{1}{4}\left(  s^{2}A\right)  \lambda^{2}\right]
=s^{a}\left[  \left(  s^{b}A\right)  \lambda_{b}\right]  +\left(  1/4\right)
s^{a}\left[  \left(  s^{2}A\right)  \lambda^{2}\right] \nonumber\\
&  =-\left(  s^{a}s^{b}A\right)  \lambda_{b}+\left(  s^{b}A\right)  \left(
s^{a}\lambda_{b}\right)  +\left(  1/4\right)  \left(  s^{2}A\right)  \left(
s^{a}\lambda^{2}\right) \nonumber\\
&  =-\left(  1/2\right)  \left(  s^{2}A\right)  \lambda^{a}-\left(  sA\right)
\left(  s^{a}\lambda\right)  +\left(  1/4\right)  \left(  s^{2}A\right)
\left(  s^{a}\lambda^{2}\right)  \label{saDeltaF}%
\end{align}
\bigskip and%
\begin{align}
s^{2}\left(  \Delta A\right)   &  =s^{2}\left[  \left(  s^{b}A\right)
\lambda_{b}+\frac{1}{4}\left(  s^{2}A\right)  \lambda^{2}\right]
=s^{2}\left[  \left(  s^{b}A\right)  \lambda_{b}\right]  +\frac{1}{4}%
s^{2}\left[  \left(  s^{2}A\right)  \lambda^{2}\right] \nonumber\\
&  =2\left(  s_{a}s^{b}A\right)  \left(  s^{a}\lambda_{b}\right)  +\left(
s^{b}A\right)  \left(  s^{2}\lambda_{b}\right)  +\frac{1}{4}\left(
s^{2}A\right)  \left(  s^{2}\lambda^{2}\right) \nonumber\\
&  =-\left(  s^{2}A\right)  \left(  s\lambda\right)  -\left(  sA\right)
\left(  s^{2}\lambda\right)  +\frac{1}{4}\left(  s^{2}A\right)  \left(
s^{2}\lambda^{2}\right)  \ . \label{s2DeltaF}%
\end{align}
Therefore, $\Delta_{\left(  1\right)  }\Delta_{\left(  2\right)  }A$ is given
by%
\begin{align}
\Delta_{\left(  1\right)  }\Delta_{\left(  2\right)  }A  &  =\left(
s^{a}\Delta_{\left(  2\right)  }A\right)  \lambda_{\left(  1\right)  a}%
+\frac{1}{4}\left(  s^{2}\Delta_{\left(  2\right)  }A\right)  \lambda_{\left(
1\right)  }^{2}\nonumber\\
&  =\left[  -\left(  1/2\right)  \left(  s^{2}A\right)  \lambda_{\left(
2\right)  }^{a}-\left(  sA\right)  \left(  s^{a}\lambda_{\left(  2\right)
}\right)  +\left(  1/4\right)  \left(  s^{2}A\right)  \left(  s^{a}%
\lambda_{\left(  2\right)  }^{2}\right)  \right]  \lambda_{\left(  1\right)
a}\nonumber\\
&  +\frac{1}{4}\left[  \left(  s^{2}A\right)  \left(  s\lambda_{\left(
2\right)  }\right)  -\left(  sA\right)  \left(  s^{2}\lambda_{\left(
2\right)  }\right)  +\frac{1}{4}\left(  s^{2}A\right)  \left(  s^{2}%
\lambda_{\left(  2\right)  }^{2}\right)  \right]  \lambda_{\left(  1\right)
}^{2}\nonumber\\
&  \equiv\left(  s^{a}A\right)  \vartheta_{\left(  1,2\right)  a}+\frac{1}%
{4}\left(  s^{2}A\right)  \theta_{\left(  1,2\right)  }\ , \label{DDeltaF}%
\end{align}
for certain functionals $\vartheta_{\left(  1,2\right)  }^{a}\left(
\Gamma\right)  $ and $\theta_{\left(  1,2\right)  }\left(  \Gamma\right)  $,
constructed from the parameters $\lambda_{\left(  j\right)  }^{a}$, for
$j=1,2$, which are generally field-dependent, $\lambda_{\left(  j\right)
}^{a}=\lambda_{\left(  j\right)  }^{a}\left(  \Gamma\right)  $,%
\begin{align}
\vartheta_{\left(  1,2\right)  }^{a}  &  =-\left(  s\lambda_{\left(  2\right)
}^{a}\right)  \lambda_{\left(  1\right)  }+\frac{1}{4}\left(  s^{2}%
\lambda_{\left(  2\right)  }^{a}\right)  \lambda_{\left(  1\right)  }%
^{2}\ ,\label{vartheta12}\\
\theta_{\left(  1,2\right)  }  &  =\left[  2\lambda_{\left(  2\right)
}-\left(  s\lambda_{\left(  2\right)  }^{2}\right)  \right]  \lambda_{\left(
1\right)  }-\left[  \left(  s\lambda_{\left(  2\right)  }\right)  -\frac{1}%
{4}\left(  s^{2}\lambda_{\left(  2\right)  }^{2}\right)  \right]
\lambda_{\left(  1\right)  }^{2}\ . \label{theta12}%
\end{align}
Hence, the commutator of finite variations reads%
\begin{equation}
\left[  \Delta_{\left(  1\right)  },\Delta_{\left(  2\right)  }\right]
A=\left(  s^{a}A\right)  \vartheta_{\left[  1,2\right]  a}+\frac{1}{4}\left(
s^{2}A\right)  \theta_{\left[  1,2\right]  }\ ,\ \ \ \vartheta_{\left[
1,2\right]  }^{a}\equiv\vartheta_{\left(  1,2\right)  }^{a}-\vartheta_{\left(
2,1\right)  }^{a}\ ,\ \ \ \theta_{\left[  1,2\right]  }\equiv\theta_{\left(
1,2\right)  }-\theta_{\left(  2,1\right)  }\ , \label{[d1d2]F}%
\end{equation}
Finally, using the identity%
\begin{equation}
\lambda_{\left(  2\right)  }\lambda_{\left(  1\right)  }-\lambda_{\left(
1\right)  }\lambda_{\left(  2\right)  }=\lambda_{\left(  2\right)  a}%
\lambda_{\left(  1\right)  }^{a}-\lambda_{\left(  1\right)  a}\lambda_{\left(
2\right)  }^{a}=\lambda_{\left(  2\right)  a}\lambda_{\left(  1\right)  }%
^{a}-\lambda_{\left(  2\right)  a}\lambda_{\left(  1\right)  }^{a}\equiv0\ ,
\label{l1l2}%
\end{equation}
we obtain%
\begin{align}
\vartheta_{\left[  1,2\right]  }^{a}=  &  \left(  s\lambda_{\left(  1\right)
}^{a}\right)  \lambda_{\left(  2\right)  }-\left(  s\lambda_{\left(  2\right)
}^{a}\right)  \lambda_{\left(  1\right)  }-\frac{1}{4}\left[  \left(
s^{2}\lambda_{\left(  1\right)  }^{a}\right)  \lambda_{\left(  2\right)  }%
^{2}-\left(  s^{2}\lambda_{\left(  2\right)  }^{a}\right)  \lambda_{\left(
1\right)  }^{2}\right]  \ ,\label{vartheta[12]}\\
\theta_{\left[  1,2\right]  }=  &  \left[  \left(  s\lambda_{\left(  1\right)
}^{2}\right)  \lambda_{\left(  2\right)  }-\left(  s\lambda_{\left(  2\right)
}^{2}\right)  \lambda_{\left(  1\right)  }\right]  +\left[  \left(
s\lambda_{\left(  1\right)  }\right)  \lambda_{\left(  2\right)  }^{2}-\left(
s\lambda_{\left(  2\right)  }\right)  \lambda_{\left(  1\right)  }^{2}\right]
\nonumber\\
&  +\frac{1}{4}\left[  \left(  s^{2}\lambda_{\left(  2\right)  }^{2}\right)
\lambda_{\left(  1\right)  }^{2}-\left(  s^{2}\lambda_{\left(  1\right)  }%
^{2}\right)  \lambda_{\left(  2\right)  }^{2}\right]  \ . \label{theta[12]}%
\end{align}
where $\vartheta_{\left[  1,2\right]  }^{a}$\textbf{, }$\theta_{\left[
1,2\right]  }$ possess the symmetry properties $\vartheta_{\left[  1,2\right]
}^{a}=-\vartheta_{\left[  2,1\right]  }^{a}$, $\theta_{\left[  1,2\right]
}=-\theta_{\left[  2,1\right]  }$. In particular, assuming $A\left(
\Gamma\right)  =\Gamma^{p}$ in (\ref{[d1d2]F}), we have
\begin{equation}
\left[  \Delta_{\left(  1\right)  },\Delta_{\left(  2\right)  }\right]
\Gamma^{p}=\left(  s^{a}\Gamma^{p}\right)  \vartheta_{\left[  1,2\right]
a}+\frac{1}{4}\left(  s^{2}\Gamma^{p}\right)  \theta_{\left[  1,2\right]  }\ .
\label{[D1D2]Ph}%
\end{equation}
In general, the commutator (\ref{[D1D2]Ph}) of finite BRST-antiBRST
transformations does not belong to the class of these transformations due to
the opposite symmetry properties of $\vartheta_{\left[  1,2\right]
a}\vartheta_{\left[  1,2\right]  }^{a}$ and $\theta_{\left[  1,2\right]  }$,%
\begin{equation}
\vartheta_{\left[  1,2\right]  a}\vartheta_{\left[  1,2\right]  }%
^{a}=\vartheta_{\left[  2,1\right]  a}\vartheta_{\left[  2,1\right]  }%
^{a}\ ,\ \ \ \theta_{\left[  1,2\right]  }=-\theta_{\left[  2,1\right]  }\ ,
\label{propfbab}%
\end{equation}
which implies that $\theta_{\left[  1,2\right]  }=\vartheta_{\left[
1,2\right]  a}\vartheta_{\left[  1,2\right]  }^{a}$ in (\ref{[D1D2]Ph}) is
possible only in the particular case $\theta_{\left[  1,2\right]  }%
=\vartheta_{\left[  1,2\right]  a}\vartheta_{\left[  1,2\right]  }^{a}=0$.
This reflects the fact that a finite nonlinear transformation has the form of
a group element, i.e., not an element of a Lie superalgebra; however, the
linear approximation $\Delta^{\mathrm{lin}}\Gamma^{p}=\left(  s^{a}\Gamma
^{p}\right)  \lambda_{a}$ to a finite transformation $\Delta\Gamma^{p}%
=\Delta^{\mathrm{lin}}\Gamma^{p}+O\left(  \lambda^{2}\right)  $ does form an
algebra; indeed, due to (\ref{[d1d2]F}), (\ref{vartheta[12]}),
(\ref{theta[12]}), we have%
\begin{equation}
\left[  \Delta_{\left(  1\right)  }^{\mathrm{lin}},\Delta_{\left(  2\right)
}^{\mathrm{lin}}\right]  A=\Delta_{\left[  1,2\right]  }^{\mathrm{lin}%
}A=\left(  s^{a}A\right)  \lambda_{\left[  1,2\right]  a}\ ,\ \ \ \lambda
_{\left[  1,2\right]  }^{a}\equiv\left(  s_{b}\lambda_{\left(  1\right)  }%
^{a}\right)  \lambda_{\left(  2\right)  }^{b}-\left(  s_{b}\lambda_{\left(
2\right)  }^{a}\right)  \lambda_{\left(  1\right)  }^{b}\ . \label{comm}%
\end{equation}
Thus, the construction of finite BRST-antiBRST transformations reduces to the
usual BRST-antiBRST transformations, $\delta\Gamma^{p}=\Delta^{\mathrm{lin}%
}\Gamma^{p}$, linear in the infinitesimal parameter $\mu_{a}=\lambda_{a}$, as
one selects the approximation that forms an algebra with respect to the commutator.

Using the above results, let us now consider an operator
$\mathcal{U}$, such that
\begin{equation}\label{UA}
\mathcal{U}A=A+\Delta A\ ,\ \ \ \mathrm{where}\ \ \ \Delta
A=\left(  s^{a}A\right) \lambda_{a}+\frac{1}{4}\left(
s^{2}A\right)  \lambda^{2}\ ,\ \ \ \Delta _{\left(  1\right)
}\Delta_{\left(  2\right)  }A=\left(  s^{a}A\right)
\vartheta_{\left(  1,2\right)  a}+\frac{1}{4}\left(  s^{2}A\right)
\theta_{\left(  1,2\right)  }\ ,
\end{equation}
and study its composition properties, namely,%
\begin{align}
\mathcal{U}_{\left(  1\right)  }\mathcal{U}_{\left(  2\right)  }A  &  = \mathcal{U}_{\left(  1\right)
}\left(  \mathcal{U}_{\left(  2\right)  }A\right)  =\mathcal{U}_{\left(  1\right)  }\left(
F+\Delta_{\left(  2\right)  }A\right)  =A+\Delta_{\left(  2\right)  }%
A+\Delta_{\left(  1\right)  }\left(  A+\Delta_{\left(  2\right)  }A\right) \\
&  =A+\Delta_{\left(  1\right)  }A+\Delta_{\left(  2\right)  }A+\Delta
_{\left(  1\right)  }\Delta_{\left(  2\right)  }A=A+s^{a}A\left[
\lambda_{\left(  1\right)  a}+\lambda_{\left(  2\right)  a}+\vartheta_{\left(
1,2\right)  a}\right]  +\frac{1}{4}s^{2}A\left[  \lambda_{\left(  1\right)
}^{2}+\lambda_{\left(  2\right)  }^{2}+\theta_{\left(  1,2\right)  }\right]
\ , \nonumber \\
\left[  \mathcal{U}_{\left(  1\right)  }, \, \mathcal{U}_{\left(  2\right)  }\right]  A  &  =\left[
\Delta_{\left(  1\right)  },\Delta_{\left(  2\right)  }\right]  A=\left(
s^{a}A\right)  \vartheta_{\left[  1,2\right]  a}+\frac{1}{4}\left(
s^{2}A\right)  \theta_{\left[  1,2\right]  }\ , \label{algUA}
\end{align}
whence follows the explicit form of the operator $\mathcal{U}$, as
well as the corresponding composition and commutator, in terms of
the operator $\overleftarrow{\mathcal{U}}$, whose action is
identical with that of $\mathcal{U}$:
\begin{align}
&  \overleftarrow{\mathcal{U}}_{\left(  1\right)
}=1+\overleftarrow{s}^{a}\lambda_{\left(  1\right)
a}+\frac{1}{4}\overleftarrow{s}^{2}\lambda_{\left(  1\right)
}^{2}\ =\ \exp\{ \overleftarrow{s}^{a}\lambda_{\left(  1\right)
a}\}\, ,\label{overT}  \\
&  \overleftarrow{\mathcal{U}}_{\left(  1,2\right)  }\equiv\overleftarrow{\mathcal{U}}_{\left(
1\right)  }\overleftarrow{\mathcal{U}}_{\left(  2\right)  }=1+\overleftarrow{s}%
^{a}\left[  \lambda_{\left(  1\right)  a}+\lambda_{\left(  2\right)
a}+\vartheta_{\left(  2,1\right)  a}\right]  +\frac{1}{4}\overleftarrow{s}%
^{2}\left[  \lambda_{\left(  1\right)  }^{2}+\lambda_{\left(  2\right)  }%
^{2}+\theta_{\left(  2,1\right)  }\right]  \ ,\label{overT12}\\
&  \left[  \overleftarrow{\mathcal{U}}_{\left(  1\right)  },\,\overleftarrow{\mathcal{U}}_{\left(
2\right)  }\right]  =\overleftarrow{\mathcal{U}}_{\left(  1,2\right)  }-\overleftarrow
{\mathcal{U}}_{\left(  2,1\right)  }=-\overleftarrow{s}^{a}\vartheta_{\left[
1,2\right]  a}-\frac{1}{4}\overleftarrow{s}^{2}\theta_{\left[  1,2\right]
}\ ,\label{overTcom}
\end{align}
with $\vartheta_{\left(  1,2\right)  a}$, $\theta_{\left(
1,2\right)  }$ and $\vartheta_{\left[  1,2\right]  a}$,
$\theta_{\left[  1,2\right]  }$ given by (\ref{vartheta12}),
(\ref{theta12}) and (\ref{vartheta[12]}), (\ref{theta[12]}).
From the above, we can see that the set of the operators $\overleftarrow{\mathcal{U}}
\equiv \overleftarrow{\mathcal{U}}\left(  \lambda \right)$ forms
an Abelian two-parametric Lie supergroup for constant odd-valued
parameters $\lambda$, $\overleftarrow{\mathcal{U}}\left( \lambda_1
\right) \overleftarrow{\mathcal{U}}\left(  \lambda_2 \right) =
\overleftarrow{\mathcal{U}}\left(  \lambda_2 \right)
\overleftarrow{\mathcal{U}}\left(  \lambda_1 \right) =
\overleftarrow{\mathcal{U}}\left(  \lambda_1 + \lambda_2 \right)$,
with the unit element $e = \overleftarrow{\mathcal{U}}\left(  0
\right)$, whereas in the case of field-dependent $\lambda$ it follows from
(\ref{overT})--(\ref{overTcom}) that the set of
$\overleftarrow{\mathcal{U}}\left(  \lambda(\Gamma) \right) $
forms a non-linear algebraic structure.

\section{Ward Identities and Gauge Dependence Problem}

\label{WIGD}\renewcommand{\theequation}{\arabic{section}.\arabic{equation}}
\setcounter{equation}{0} We can now apply finite BRST-antiBRST transformations
to derive modified Ward (Slavnov--Taylor) identities and to study the problem
of gauge-dependence for the generating functional of Green's functions
(\ref{ZPI}). As compared to the partition function $Z_{\Phi}$ in
(\ref{zphizphi1}), the functional $Z_{\Phi}(I)$ in the presence of external
sources $I_{p}(t)$ should depend on a choice of the gauge Boson ${\Phi}$;
however, in view of the equivalence theorem \cite{equiv}, this dependence is
highly structured, so that physical quantities cannot \textquotedblleft
feel\textquotedblright\ gauge dependence.

Using (\ref{finBRSTantiBRST2}), the relation (\ref{DelFfrule}) for functionals,
and the relations (\ref{varSH}), (\ref{pottrans}) for the action $S_{H,\Phi}$,
we have%
\begin{equation}
S_{H,\Phi}(\check{\Gamma})=S_{H,\Phi}\left(  {\Gamma}\right)  \left(
1+\overleftarrow{s}^{a}\lambda_{a}+\frac{1}{4}\overleftarrow{s}^{2}\lambda
^{2}\right)  \ , \label{transruleS}%
\end{equation}
where the operators $\overleftarrow{s}^{a}$ act in accordance with (\ref{y}).
Then, using (\ref{transruleS}) and (\ref{pottrans}), we obtain the formula%
\begin{equation}
S_{H,\Phi}(\check{\Gamma})=S_{H,\Phi}\left(  {\Gamma}\right)  +\frac{1}%
{2}\left.  \left[  \left(  \Gamma^{p}\partial_{p}\Omega^{a}-2\Omega
^{a}\right)  \lambda_{a}+\frac{1}{4}\Gamma^{p}s_{a}(\partial_{p}\Omega
^{a})\lambda^{2}\right]  \right\vert _{t_{\mathrm{in}}}^{t_{\mathrm{out}}}\ .
\label{transruleS2}%
\end{equation}
In terms of $\overleftarrow{s}^{a}$, the functional Jacobian (\ref{superJ1})
has the form%
\begin{equation}
\exp(\Im)=\left[  1-\frac{1}{2}\left(  \int dt\,\Lambda(t)\right)
\overleftarrow{s}^{2}\right]  ^{-2}. \label{funcJacobian}%
\end{equation}

Let us subject (\ref{ZPI}) to a field-dependent BRST-antiBRST transformation
of trajectories (\ref{deffin_traj}). Then, the relation (\ref{funcJacobian})
for the Jacobian and the properties
(\ref{varSH}), (\ref{transruleS2}) of gauge invariance for the action
allow one to obtain a \emph{modified Ward}
(\emph{Slavnov--Taylor}) \emph{identity:}%
\begin{align}
&  \left\langle \left\{  1+\frac{i}{\hbar}\int dtI_{p}(t)\Gamma^{p}(t)\left(
\overleftarrow{s}^{a}\lambda_{a}(\Lambda)+\frac{1}{4}\overleftarrow{s}%
^{2}\lambda^{2}(\Lambda)\right)  -\frac{1}{4}\left(  \frac{i}{\hbar}\right)
{}^{2}\int dt\ dt^{\prime}\ I_{p}(t)\Gamma^{p}(t)\overleftarrow{s}^{a}%
I_{q}(t^{\prime})\Gamma^{q}(t^{\prime})\overleftarrow{s}_{a}\lambda
^{2}(\Lambda)\right\}  \right. \nonumber\\
&  \quad\left.  \times\left\{  1-\frac{1}{2}\left[  \int dt\Lambda(t)\right]
\overleftarrow{s}^{2}\right\}  {}^{-2}\right\rangle _{\Phi,I}=1, \label{mWI}%
\end{align}
where the symbol \textquotedblleft$\langle\mathcal{O}\rangle_{\Phi,I}$
\textquotedblright\ for any quantity $\mathcal{O}=\mathcal{O}(\Gamma)$
denotes a source-dependent average expectation value corresponding to
a gauge $\Phi(\Gamma)$, namely,%
\begin{equation}
\left\langle \mathcal{O}\right\rangle _{\Phi,I}=Z_{\Phi}^{-1}(I)\int
d\Gamma\ \mathcal{O}\exp\left\{  \frac{i}{\hbar}\left[  S_{H,\Phi}%
(\Gamma)+\int dt\,I(t)\Gamma(t)\right]  \right\}  \ ,\ \ \ \mathrm{with\ \ \ }%
\left\langle 1\right\rangle _{\Phi,I}=1\ . \label{aexv}%
\end{equation}
In (\ref{mWI}), both $\Lambda(\Gamma)$ and $I_{p}(t)$ are arbitrary, so that,
due to the explicit presence of $\Lambda(\Gamma)$ [which implies $\lambda
_{a}(\Lambda)$], the modified Ward identity implicitly depends on a choice of
the gauge Bosonic function $\Phi(\Gamma)$ for non-vanishing $I_{p}(t)$,
according to (\ref{solcompeq2}), (\ref{funcdeplafin}). Thus, the corresponding
Ward identities for Green's functions obtained by differentiating (\ref{mWI})
with respect to sources contain functionals $\lambda_{a}(\Lambda)$ and their
derivatives [implicitly, $\Phi(\Gamma)$] as weight functionals, as compared
to the usual Ward identities for constant $\lambda_{a}$. Indeed, for
$\lambda_{a}=\mathrm{const}$ the identity (\ref{mWI}) implies two independent
Ward identities at the first degree in powers of $\lambda_{a}$,%
\[
\left\langle \int dt\ I_{p}(t)\Gamma^{p}(t)\overleftarrow{s}^{a}\right\rangle
_{\Phi,I}=0\ ,
\]
which are identical with those of (\ref{WIham}), as well as a new Ward
identity at the second degree in powers of $\lambda_{a}$,%
\[
\left\langle \int dt\,I_{p}(t)\Gamma^{p}(t)\left[  \overleftarrow{s}%
^{2}-\overleftarrow{s}^{a}\left(  \frac{i}{\hbar}\right)  \int
dt^{\prime
}\ I_{q}(t^{\prime})\left(  \Gamma^{q}(t^{\prime})\overleftarrow{s}%
_{a}\right)  \right]  \right\rangle _{\Phi,I}=0\ .
\]
Substituting, instead of $\lambda_{a}(\Lambda)$ [and
$\Lambda(\Gamma)$] in (\ref{mWI}), the solution
(\ref{funcdeplafin}) [(\ref{solcompeq2})] of the compensation
equation (\ref{eqexpl}), we obtain, according to the study
of Section~\ref{compEq}, the following relation:
\begin{align}
Z_{\Phi+\Delta\Phi}(I)  &  =Z_{\Phi}(I)\left\{  1+\left\langle\frac{i}{\hbar}\int
dt\ I_{p}(t)\left[  (s^{a}\Gamma^{p}(t))\lambda_{a}\left(  \Gamma|-\Delta
{\Phi}\right)  +\frac{1}{4}(s^{2}\Gamma^{p}(t))\lambda^{2}\left(
\Gamma|-\Delta{\Phi}\right)  \right]  \right.\right. \nonumber\\
&  -\left.\left.  (-1)^{\varepsilon_{q}}\left(  \frac{i}{2\hbar}\right)  ^{2}\int
dt\ dt^{\prime}I_{q}(t^{\prime})I_{p}(t)(s^{a}\Gamma^{p}(t))(s_{a}\Gamma
^{q}(t^{\prime}))\lambda^{2}\left(  \Gamma|-\Delta{\Phi}\right)\right\rangle  \right\}  \;,
\label{GDI}%
\end{align}
which extends the result (\ref{zphizphi1}) to non-vanishing external sources
$I_{p}(t)$.

Following \cite{BLT1h}, let us now enlarge the generating
functional $Z_{\Phi}(I)$ to an extended
generating functional of Green's functions $Z_{\Phi}(I,\Gamma^{\ast}%
,\overline{\Gamma})$ by adding to the action $S_{H,\Phi}$ some new
terms with external sources (antifields) $\Gamma_{pa}^{\ast}(t)$
for $a=1,2$ and $\overline{\Gamma}_{p}(t)$,
$\varepsilon(\Gamma_{pa}^{\ast})+1$ =
$\varepsilon(\overline{\Gamma}_{p})$ = $\varepsilon_{p}$\,,
multiplied by the respective BRST-antiBRST variations $(s^{a}\Gamma^{p})(t)$
and their commutator $(s^{2}\Gamma^{p})(t)$, namely,
\begin{equation}
Z_{\Phi}(I,\Gamma^{\ast},\overline{\Gamma})=\int d\Gamma\ \exp\left\{
\frac{i}{\hbar}\left[  S_{H,\Phi}(\Gamma)+\int dt\left(  \Gamma_{pa}^{\ast
}s^{a}\Gamma^{p}-\frac{1}{2}\overline{\Gamma}_{p}s^{2}\Gamma^{p}%
+I\Gamma\right)  \right]  \right\}  \ ,\ \ \mathrm{for}\ \ Z_{\Phi
}(I,0,0)=Z_{\Phi}(I). \label{ZPIext}%
\end{equation}
If we make in (\ref{ZPIext}) a change of variables (trajectories) in the
extended space $(\Gamma^{p},\Gamma_{pa}^{\ast},\overline{\Gamma}_{p})$,%
\begin{align}
&  \Gamma^{p}(t)\rightarrow\check{\Gamma}^{p}(t)=\Gamma^{p}\left(
t\right)  \left(  1+\overleftarrow{s}^{a}\lambda_{a}+\frac{1}{4}%
\overleftarrow{s}^{2}\lambda^{2}\right)  \ ,\nonumber\\
&  \Gamma_{pa}^{\ast}(t)\rightarrow\check{\Gamma}{}_{pa}^{\ast}(t)=\Gamma
_{pa}^{\ast}(t)\ ,\label{extBRSTantiBRST}\\
&  \overline{\Gamma}_{p}(t)\rightarrow\check{\overline{\Gamma}}_{p}%
(t)=\overline{\Gamma}_{p}(t)-\varepsilon^{ab}\lambda_{a}\Gamma_{pb}^{\ast
}(t)\ ,\nonumber
\end{align}
for $I_{p}=0$ with finite constant parameters $\lambda_{a}$, we find that the
integrand in (\ref{ZPIext}) remains the same, in view of $\overleftarrow{s}%
_{a}\overleftarrow{s}_{b}\overleftarrow{s}_{c}\equiv0$ and due to
$\Delta\left(  \Gamma_{pa}^{\ast}s^{a}\Gamma^{p}+\frac{1}{2}\overline{\Gamma
}_{p}s^{2}\Gamma^{p}\right)  =0$, which implies that the transformations
(\ref{extBRSTantiBRST}) are \emph{extended BRST-antiBRST transformations} for
the functional $Z_{\Phi}(I,\Gamma^{\ast},\overline{\Gamma})$.

Making in (\ref{ZPIext}) a change of variables, which corresponds only to
BRST-antiBRST transformations $\Gamma^{p}(t)\rightarrow\check{\Gamma}^{p}(t)$
with an arbitrary functional $\lambda_{a}(\Gamma)=\int dt\Lambda
(t)\overleftarrow{s}_{a}$ from (\ref{funcdepla}), we obtain a \emph{modified
Ward identity} for $Z_{\Phi}(I,\Gamma^{\ast},\overline{\Gamma})$:
\begin{align}
&  \hspace{-1.5em}\left\langle \left\{  1+\frac{i}{\hbar}\int
dt\left[
I_{p}\left(  \Gamma^{p}\overleftarrow{s}^{a}\lambda_{a}(\Lambda)+\frac{1}%
{4}\Gamma^{p}\overleftarrow{s}^{2}\lambda^{2}(\Lambda)\right)  +\frac{1}%
{2}\varepsilon^{ab}\Gamma_{pb}^{\ast}(\Gamma^{p}\overleftarrow{s}^{2}%
)\lambda_{a}\right]  +\frac{\varepsilon_{ab}}{4}\left(
\frac{i}{\hbar
}\right)  {}^{2}\int dt\ \left[  I_{p}(\Gamma^{p}\overleftarrow{s}%
^{a})\phantom{\int}\right.  \right.  \right.  \nonumber\\
&  \ \hspace{-1.5em}\left.  \left.  \left.  +\frac{1}{2}\varepsilon^{ac}%
\Gamma_{pc}^{\ast}(\Gamma^{p}\overleftarrow{s}^{2})\right]  \int
dt^{\prime }\left[
I_{q}(\Gamma^{q}\overleftarrow{s}^{b})+\frac{1}{2}\varepsilon
^{bd}\Gamma_{qd}^{\ast}(\Gamma^{q}\overleftarrow{s}^{2})\right]
\lambda ^{2}(\Lambda)\right\}  \left\{  1-\frac{1}{2}\left[  \int
dt\Lambda(t)\right] \overleftarrow{s}^{2}\right\}
{}^{-2}\right\rangle _{\Phi,I,\Gamma^{\ast
},\overline{\Gamma}}=1\ ,\label{mWIext}%
\end{align}
where the symbol \textquotedblleft$\langle\mathcal{O}\rangle_{\Phi
,I,\Gamma^{\ast},\overline{\Gamma}}$\textquotedblright\ for any $\mathcal{O}%
=\mathcal{O}(\Gamma)$ stands for a source-dependent average expectation value
for a gauge $\Phi(\Gamma)$ in the presence of the antifields $\Gamma
_{pa}^{\ast},\overline{\Gamma}_{p}$\,, namely,
\begin{align}
\left\langle \mathcal{O}\right\rangle _{\Phi,I,\Gamma^{\ast},\overline{\Gamma
}}  &  =Z_{\Phi}^{-1}(I,\Gamma^{\ast},\overline{\Gamma})\int d\Gamma
\ \mathcal{O}\exp\left\{  \frac{i}{\hbar}\left[  S_{H,\Phi}(\Gamma
,\Gamma^{\ast},\overline{\Gamma})+\int dt\ I(t)\Gamma(t)\right]  \right\}
,\label{aexvext}\\
&  \mathrm{with}\ \ \ S_{H,\Phi}(\Gamma,\Gamma^{\ast},\overline{\Gamma
})=S_{H,\Phi}(\Gamma)+\int dt\left(  \Gamma_{pa}^{\ast}s^{a}\Gamma^{p}%
-\frac{1}{2}\overline{\Gamma}_{p}s^{2}\Gamma^{p}\right)  \ .\nonumber
\end{align}
We can see that the difference of (\ref{mWI}) and (\ref{mWIext})
is in the definitions (\ref{aexv}) and (\ref{aexvext}),
as well as in the presence of the terms
proportional to $(1/2)\varepsilon^{ab}\Gamma_{pb}^{\ast}(\Gamma
^{p}\overleftarrow{s}^{2})$ at the first and second degrees in powers of
$\lambda_{a}$, except for the Jacobian.

For constant parameters $\lambda_{a}$, we deduce from (\ref{mWIext})
\begin{align}
\left\langle \int dt\ \left[  I_{p}(t)\Gamma^{p}(t)\overleftarrow{s}^{a}%
+\frac{1}{2}\varepsilon^{ab}\Gamma_{pb}^{\ast}(t)(\Gamma^{p}(t)\overleftarrow
{s}^{2})\right]  \right\rangle _{\Phi,I,\Gamma^{\ast},\overline{\Gamma}%
}=0\ , \label{WIham1ext}%
\end{align}
as well as a new Ward identity at the second degree in powers of $\lambda_{a}%
$:%
\begin{align}
&  \left\langle \int dt\ I_{p}(t)\Gamma^{p}(t)\overleftarrow{s}^{2}%
+\varepsilon_{ab}\left(  \frac{i}{\hbar}\right)  \int dt\ \left[  I_{p}%
\Gamma^{p}\overleftarrow{s}^{a}+\frac{1}{2}\varepsilon^{ac}\Gamma_{pc}^{\ast
}(\Gamma^{p}\overleftarrow{s}^{2})\right]  (t)\right. \nonumber\\
&  \times\left.  \int dt^{\prime}\ \left[  I_{q}(\Gamma^{q}\overleftarrow
{s}^{b}+\frac{1}{2}\varepsilon^{bd}\Gamma_{qd}^{\ast}(\Gamma^{q}%
\overleftarrow{s}^{2})\right]  (t^{\prime})\right\rangle _{\Phi,I,\Gamma
^{\ast},\overline{\Gamma}}=0\ . \label{WIham2ext}%
\end{align}
The respective identities (\ref{WIham1ext}) and (\ref{WIham2ext}) may be
represented as%
\begin{align}
\int dt\ \left[
I_{p}(t)\frac{\overrightarrow{\delta}}{\delta\Gamma
_{pa}^{\ast}(t)}-\varepsilon^{ab}\Gamma_{pb}^{\ast}(t)\frac{\overrightarrow
{\delta}}{\delta\overline{\Gamma}{}_{p}(t)}\right]  \ln Z_{\Phi}%
(I,\Gamma^{\ast},\overline{\Gamma})=0\ , \label{WIham1exteq}%
\end{align}
and%
\[
\varepsilon_{ab}\int dt\ dt^{\prime}\ \left[
I_{p}(t)\frac{\overrightarrow
{\delta}}{\delta\Gamma_{pa}^{\ast}(t)}-\varepsilon^{ac}\Gamma_{pc}^{\ast
}(t)\frac{\overrightarrow{\delta}}{\delta\overline{\Gamma}{}_{p}(t)}\right]
\left[
I_{q}(t^{\prime})\frac{\overrightarrow{\delta}}{\delta\Gamma
_{qb}^{\ast}(t^{\prime})}-\varepsilon^{bd}\Gamma_{qd}^{\ast}(t^{\prime}%
)\frac{\overrightarrow{\delta}}{\delta\overline{\Gamma}{}_{q}(t^{\prime}%
)}\right]  \ln Z_{\Phi}(I,\Gamma^{\ast},\overline{\Gamma})=0\ ,
\]
being a differential consequence of (\ref{WIham1exteq}) which
follows from applying to the latter the operators
\[
\int dt^{\prime}\left[  I_{q}(t^{\prime})\frac{\overrightarrow{\delta}}%
{\delta\Gamma_{qb}^{\ast}(t^{\prime})}-\varepsilon^{bd}\Gamma_{qd}^{\ast}(t^{\prime})%
\frac{\overrightarrow{\delta}}{\delta\overline{\Gamma}{}_{q}(t^{\prime}%
)}\right] .
\]

Let us consider the functional $S(\Gamma,\Gamma^{\ast},\overline{\Gamma})$
being a functional Legendre transform of $\ln Z_{\Phi}(I,\Gamma^{\ast
},\overline{\Gamma})$ with respect to the sources $I_{p}(t)$:%
\begin{align}
&  \Gamma^{p}=\frac{\hbar}{i}\frac{\overrightarrow{\delta}}{\delta I_{p}}\ln
Z_{\Phi}(I,\Gamma^{\ast},\overline{\Gamma})\ ,\label{avfields}\\
&  S(\Gamma,\Gamma^{\ast},\overline{\Gamma})=\frac{\hbar}{i}\ln Z_{\Phi
}(I,\Gamma^{\ast},\overline{\Gamma})-\int dt\ I_{p}(t)\Gamma^{p}%
(t)\ ,\label{avfields1}\\
&  \mathrm{where}\ \ I_{p}(t)=-S(\Gamma,\Gamma^{\ast},\overline{\Gamma}%
)\frac{\overleftarrow{\delta}}{\delta\Gamma^{p}(t)}\ . \label{avfields2}%
\end{align}
From (\ref{WIham1exteq})--(\ref{avfields2}), we obtain an
$\mathrm{Sp}\left(  2\right)  $-doublet of independent Ward
identities for $S(\Gamma
,\Gamma^{\ast},\overline{\Gamma})$,%
\begin{equation}
\frac{1}{2}\left(  S,S\right)  ^{a}+V^{a}S=0\ , \label{meq}%
\end{equation}
in terms of the $\mathrm{Sp}\left(  2\right)  $-doublets of
extended antibrackets and operators $V^{a}$ known from the
$\mathrm{Sp}\left( 2\right)  $-covariant Lagrangian quantization
\cite{BLT1,BLT2} for gauge theories:
\begin{equation}
\left(  F,G\right)  ^{a}=\int dt\ F\left(  \frac{\overleftarrow{\delta}%
}{\delta\Gamma^{p}(t)}\frac{\overrightarrow{\delta}}{\delta\Gamma_{pa}^{\ast
}(t)}-\frac{\overleftarrow{\delta}}{\delta\Gamma_{pa}^{\ast}(t)}%
\frac{\overrightarrow{\delta}}{\delta\Gamma^{p}(t)}\right)  G\ ,\ \ \ V^{a}%
=\varepsilon^{ab}\int dt\ \Gamma_{pb}^{\ast}(t)\frac{\overrightarrow{\delta}%
}{\delta\overline{\Gamma}{}_{p}(t)}\ . \label{eantibr}%
\end{equation}

\section{Relating Different Hamiltonian Gauges in Yang--Mills Theories}

\label{YMgauges}%
\renewcommand{\theequation}{\arabic{section}.\arabic{equation}} \setcounter{equation}{0}

In this section, we examine the Yang--Mills theory, given by the Lagrangian
action%
\begin{equation}
S_{0}(A)=-\frac{1}{4}\int d^{D}x\ F_{\mu\nu}^{\mathsf{u}}F^{\mathsf{u}\mu\nu
}\ ,\ \ \ \mathrm{for}\ \ \ F_{\mu\nu}^{\mathsf{u}}=\partial_{\mu}A_{\nu
}^{\mathsf{u}}-\partial_{\nu}A_{\mu}^{\mathsf{u}}+f^{\mathsf{uvw}}A_{\mu
}^{\mathsf{w}}A_{\nu}^{\mathsf{v}}\,, \label{4.1}%
\end{equation}
with the Lorentz indices $\mu,\nu=0,1,\ldots,D{-}1$, the metric tensor
$\eta_{\mu\nu}=\mathrm{diag}(-,+,\ldots,+)$, and the totally antisymmetric
$su(N)$ structure constants $f^{\mathsf{uvw}}$ for $\mathsf{u},\mathsf{v}%
,\mathsf{w}=1,\ldots,N^{2}-1$.

Let us consider the given gauge theory in the BRST-antiBRST generalized
Hamiltonian quantization \cite{BLT1h,BLT2h}. To this end, note that the
corresponding dynamical system is described in the initial phase space $\eta$
[$x^{\mu}=\left(  t,\boldsymbol{x}\right)  $, $t=x^{0}$, $\boldsymbol{x}%
=\left(  x^{1},\ldots,x^{D-1}\right)  $, with the spatial indices being
denoted as $k$, $l$: $\mu=(0,k)$]%
\[
\eta=(p_{i},q^{i})=(\Pi_{k}^{\mathsf{u}},A^{\mathsf{u}k}%
)\ ,\ \ \ i=(k,\mathsf{u},\boldsymbol{x})
\]
by the classical Hamiltonian $H_{0}\left(  \eta\right)  $%
\begin{equation}
H_{0}=\int d\boldsymbol{x}\left(  -\frac{1}{2}\Pi_{k}^{\mathsf{u}}%
\Pi^{\mathsf{u}k}+\frac{1}{4}F_{kl}^{\mathsf{u}}F^{\mathsf{u}kl}\right)
\label{H0}%
\end{equation}
and by the set of linearly-independent constraints $T_{\alpha}\left(
\eta\right)  $, $\alpha=($\textsf{$u$}$,\boldsymbol{x})$,
\begin{equation}
T_{\alpha}\equiv T^{\mathsf{u}}=D_{k}^{\mathsf{uv}}\Pi^{\mathsf{v}%
k}\ ,\ \ \ D_{k}^{\mathsf{uv}}=\delta^{\mathsf{uv}}\partial_{k}%
+f^{\mathsf{uwv}}A_{k}^{\mathsf{w}}\ , \label{T0Dk}%
\end{equation}
with the following involution relations:%
\begin{equation}
\{T^{\mathsf{u}}\left(  t\right)  ,H_{0}\left(  t\right)
\}=0\ ,\ \ \ \{T^{\mathsf{u}}(t,\boldsymbol{x}),T^{\mathsf{v}}%
(t,\boldsymbol{y})\}=\int d\boldsymbol{z}\;f^{\mathsf{uvw}}T^{\mathsf{w}%
}(t,\boldsymbol{z})\delta(\boldsymbol{x}-\boldsymbol{z})\delta(\boldsymbol{y}%
-\boldsymbol{z})\ . \label{explVU}%
\end{equation}
Hence, the structure coefficients $V_{\alpha}^{\beta}$, $U_{\alpha\beta
}^{\gamma}$ arising in (\ref{invrel}) are given by [$\alpha=($\textsf{$u$%
}$,\boldsymbol{x})$, $\beta=($\textsf{$v$}$,\boldsymbol{y})$, $\gamma
=($\textsf{$w$}$,\boldsymbol{z})$]%
\[
V_{\alpha}^{\beta}=0\ ,\ \ \ U_{\alpha\beta}^{\gamma}\equiv U^{\mathsf{uvw}%
}=f^{\mathsf{uvw}}\delta(\boldsymbol{x}-\boldsymbol{z})\delta(\boldsymbol{y}%
-\boldsymbol{z})\ .
\]
The extended phase space $\Gamma$ of the given irreducible dynamical system
has the form%
\[
\Gamma=(P_{A},Q^{A})=(\Pi_{k}^{\mathsf{u}},A^{\mathsf{u}k},\mathcal{P}%
_{a}^{\mathsf{u}},C^{\mathsf{u}a},\lambda^{\mathsf{u}},\pi^{\mathsf{u}})\ ,
\]
where the Grassmann parity and the ghost number of the variables $\Gamma$ read
as follows:%
\[
\varepsilon(\Gamma)=(0,0,1,1,0,0)\ ,\ \ \ \mathrm{gh}(\Gamma)=(0,0,\left(
-1\right)  ^{a},\left(  -1\right)  ^{a+1},0,0)\ .
\]
The explicit form of the structure coefficients and of the extended phase
space $\Gamma$ allows one to construct explicit solutions \cite{Spiridonov,RML}
to the generating equations (\ref{HOmega}) with the boundary conditions
(\ref{bcond}) for the functions $\mathcal{H}$, $\Omega^{a}$, namely,%
\begin{align}
\mathcal{H}  &  =H_{0}\ ,\nonumber\\
\Omega^{a}  &  =\int d\boldsymbol{x}\left(  C^{\mathsf{u}a}D_{k}^{\mathsf{uv}%
}\Pi^{\mathsf{v}k}+\varepsilon^{ab}\mathcal{P}_{b}^{\mathsf{u}}\pi
^{\mathsf{u}}+\frac{1}{2}\mathcal{P}_{b}^{\mathsf{w}}f^{\mathsf{wvu}%
}C^{\mathsf{u}a}C^{\mathsf{v}b}\right. \nonumber\\
&  \left.  -\frac{1}{2}\lambda^{\mathsf{w}}f^{\mathsf{wvu}}C^{\mathsf{u}a}%
\pi^{\mathsf{v}}-\frac{1}{12}\lambda^{\mathsf{w}}f^{\mathsf{wvu}%
}f^{\mathsf{uts}}C^{\mathsf{s}a}C^{\mathsf{t}b}C^{\mathsf{v}c}\varepsilon
_{bc}\right)  \ . \label{4.9}%
\end{align}
Using (\ref{4.9}), let us consider the generating functional of Green's
functions $Z(I)$, given by (\ref{ZPI}). To do so, we choose the following
Bosonic gauge function $\Phi$ in the relation (\ref{Hphi}) for the unitarizing
Hamiltonian $H_{\Phi}$:%
\begin{equation}
\Phi=\int d\boldsymbol{x}\left(  -\frac{\alpha}{2}A_{k}^{\mathsf{u}%
}A^{\mathsf{u}k}+\frac{1}{2\alpha}\lambda^{\mathsf{u}}\lambda^{\mathsf{u}%
}-\frac{\beta}{2}\varepsilon_{ab}C^{\mathsf{u}a}C^{\mathsf{u}b}\right)  \,.
\label{Phi}%
\end{equation}
The unitarizing Hamiltonian $H_{\Phi}$ in (\ref{expval}) has the form%
\[
H_{\Phi}(t)=\int d\boldsymbol{x}\left(  -\frac{1}{2}\Pi_{k}^{\mathsf{u}}%
\Pi^{\mathsf{u}k}+\frac{1}{4}F_{kl}^{\mathsf{u}}F^{\mathsf{u}kl}\right)
+\frac{1}{2}\varepsilon_{ab}\left\{  \left\{  \Phi,\Omega^{a}\right\}
,\Omega^{b}\right\}  \ ,
\]
where%
\begin{align}
\frac{1}{2}\varepsilon_{ab}\left\{  \left\{  \Phi,\Omega^{a}\right\}
,\Omega^{b}\right\}   &  =\int d\boldsymbol{x}\left[  -\alpha\left(  \frac
{1}{2}\varepsilon_{ab}C^{\mathsf{u}b}D_{k}^{\mathsf{uv}}\left(  \partial
^{k}C^{\mathsf{u}a}\right)  +\partial_{k}A^{\mathsf{u}k}\pi^{\mathsf{u}%
}\right)  \right. \nonumber\\
&  +\frac{1}{2\alpha}\left(  \varepsilon^{ab}\mathcal{P}_{a}^{\mathsf{u}%
}\mathcal{P}_{b}^{\mathsf{u}}+2\lambda^{\mathsf{u}}\mathcal{P}_{a}%
^{\mathsf{v}}f^{\mathsf{vuw}}C^{\mathsf{w}a}-2\lambda^{\mathsf{u}}%
D_{k}^{\mathsf{uv}}\Pi^{\mathsf{v}k}-\frac{1}{4}\lambda^{\mathsf{u}}%
\lambda^{\mathsf{v}}f^{\mathsf{vtw}}f^{\mathsf{wsu}}C^{\mathsf{s}%
c}C^{\mathsf{t}d}\varepsilon_{dc}\right) \nonumber\\
&  +\left.  \beta\left(  \pi^{\mathsf{u}}\pi^{\mathsf{u}}-\frac{1}%
{24}f^{\mathsf{vuw}}f^{\mathsf{wts}}C^{\mathsf{s}a}C^{\mathsf{t}%
c}C^{\mathsf{u}b}C^{\mathsf{v}d}\varepsilon_{ab}\varepsilon_{cd}\right)
\right]  \ . \label{where}%
\end{align}
Integrating in the functional integral (\ref{ZPI}) over the momenta $\Pi
_{k}^{\mathsf{u}}$, $\mathcal{P}_{a}^{\mathsf{u}}$ and assuming the
corresponding sources to be equal to zero, we obtain, with allowance made for
the notation \cite{RML}
\begin{equation}
A_{0}^{\mathsf{u}}\equiv\alpha^{-1}\lambda^{\mathsf{u}}\ ,\ \ \ B^{\mathsf{u}%
}\equiv\pi^{\mathsf{u}}\ , \label{identif}%
\end{equation}
the following representation for the generating functional of Green's
functions (\ref{ZPI}) in the space of fields $\phi^{A}\left(  t,\boldsymbol{x}\right)=\left(  A^{\mathsf{u}%
\mu},B^{\mathsf{u}},C^{\mathsf{u}a}\right)\left(  t,\boldsymbol{x}\right)  $ with the corresponding sources
$J_{A}\left(  t,\boldsymbol{x}\right)$:
\begin{equation}
Z\left(  J\right)  =\int d\phi\exp\left\{  \frac{i}{\hbar}\left[  S_{0}%
(\phi)+S_{\mathrm{gf}}\left(  A,B\right)  +S_{\mathrm{gh}}\left(  A,C\right)
+S_{\mathrm{add}}\left(  C\right)  +\int dt\ J_{A}\left(t\right)\phi^{A}\left(t\right)\right]  \right\}
\,, \label{S(A,B,C)}%
\end{equation}
where the gauge-fixing term $S_{\mathrm{gf}}$, the ghost term $S_{\mathrm{gh}%
}$, and the interaction term $S_{\mathrm{add}}$, quartic in $C^{\mathsf{u}a}$,
are given by%
\begin{align}
S_{\mathrm{gf}}  &  =\int d^{D}x\ \left[  \alpha\left(  \partial^{\mu}A_{\mu
}^{\mathsf{u}}\right)  -\beta B^{\mathsf{u}}\right]  B^{\mathsf{u}%
}\,,\,\,\,S_{\mathrm{gh}}=\frac{\alpha}{2}\int d^{D}x\ \left(  \partial^{\mu
}C^{\mathsf{u}a}\right)  D_{\mu}^{\mathsf{uv}}C^{\mathsf{v}b}\varepsilon
_{ab}\ ,\label{Sgh}\\
S_{\mathrm{add}}  &  =\frac{\beta}{24}\int d^{D}x\ \ f^{\mathsf{vuw}%
}f^{\mathsf{wts}}C^{\mathsf{s}a}C^{\mathsf{t}c}C^{\mathsf{u}b}C^{\mathsf{v}%
d}\varepsilon_{ab}\varepsilon_{cd}\,, \label{Sadd}%
\end{align}
which differs from the result of \cite{RML}, corresponding to the choice
$\beta=0$ in (\ref{Phi}), by the presence of the term quadratic in
$B^{\mathsf{u}}$ and the term quartic in $C^{\mathsf{u}a}$. The result of
integration (\ref{S(A,B,C)}) is identical with\ the generating functional of
Green's functions\ recently obtained in \cite{MRnew} by the Lagrangian
BRST-antiBRST quantization of the Yang--Mills theory. This coincidence
establishes the unitarity of the $S$-matrix in the Lagrangian approach
of \cite{MRnew}.

Let us examine the choice of $\alpha$, $\beta$ leading to $R_{\xi}$-like
gauges. Namely, in view of the contribution $S_{\mathrm{gf}}$%
\begin{equation}
S_{\mathrm{gf}}=\int d^{D}x\ \left[  \alpha\left(  \partial^{\mathsf{u}}%
A_{\mu}^{\mathsf{u}}\right)  -\beta B^{\mathsf{u}}\right]  B^{\mathsf{u}}\ ,
\label{Sgfxi}%
\end{equation}
we impose the conditions%
\begin{equation}
\alpha=1\ ,\ \ \ \beta=-\frac{\xi}{2}\ . \label{coeffab}%
\end{equation}
Thus, the gauge-fixing function $\Phi_{\left(  \xi\right)  }=\Phi_{\left(
\xi\right)  }\left(  \Gamma\right)  $ corresponding to an $R_{\xi}$-like gauge
can be chosen as%
\begin{align}
\Phi_{\left(  \xi\right)  }  &  =\frac{1}{2}\int d\boldsymbol{x}\left(
-A_{k}^{\mathsf{u}}A^{\mathsf{u}k}+\lambda^{\mathsf{u}}\lambda^{\mathsf{u}%
}+\frac{\xi}{2}\varepsilon_{ab}C^{\mathsf{u}a}C^{\mathsf{u}b}\right)
\,,\,\,\,\mathrm{so}\,\,\mathrm{that}\label{Fxi}\\
\Phi_{\left(  0\right)  }  &  =\frac{1}{2}\int d\boldsymbol{x}\left(
-A_{k}^{\mathsf{u}}A^{\mathsf{u}k}+\lambda^{\mathsf{u}}\lambda^{\mathsf{u}%
}\right)  \ \ \ \mathrm{and}\ \ \ \Phi_{\left(  1\right)  }\ =\ \frac{1}%
{2}\int d\boldsymbol{x}\left(  -A_{k}^{\mathsf{u}}A^{\mathsf{u}k}%
+\lambda^{\mathsf{u}}\lambda^{\mathsf{u}}+\frac{1}{2}\varepsilon
_{ab}C^{\mathsf{u}a}C^{\mathsf{u}b}\right)  \ , \label{F01}%
\end{align}
where the gauge-fixing function $\Phi_{\left(  0\right)  }$ induces the
contribution $S_{\mathrm{gf}}\left(  A,B\right)  $ to the quantum action that
arises in the case of the Landau gauge $\partial^{\mu}A_{\mu}^{\mathsf{u}}=0$
for $(\alpha,\beta)=(1,0)$ in (\ref{Sgfxi}), whereas the function
$\Phi_{\left(  1\right)  }\left(  A,C\right)  $ corresponds to the Feynman
(covariant) gauge $\partial^{\mu}A_{\mu}^{\mathsf{u}}+\left(  1/2\right)
B^{\mathsf{u}}=0$ for $(\alpha,\beta)=(1,-1/2)$ in (\ref{Sgfxi}).

Let us find the parameters $\lambda_{a}=\int dt\ s_{a}\Lambda$ of a finite
field-dependent BRST-antiBRST transformation that connects an $R_{\xi}$ gauge
with an $R_{\xi+\Delta\xi}$ gauge:%
\begin{equation}
\Delta\Phi_{\left(  \xi\right)  }=\Phi_{\left(  \xi+\Delta\xi\right)  }%
-\Phi_{\left(  \xi\right)  }=\frac{\Delta\xi}{4}\varepsilon_{ab}\int
d\boldsymbol{x}\ C^{\mathsf{u}a}C^{\mathsf{u}b}\ . \label{DFxi}%
\end{equation}
Choosing the solution (\ref{solcompeq2}) of the compensation equation
(\ref{eqexpl}) according to the choice $\Delta\Phi=-\Delta\Phi_{\left(
\xi\right)  }$, we have%
\begin{equation}
\Lambda(\Gamma|-\Delta\Phi_{\left(  \xi\right)  })=-\frac{1}{2i\hbar
}g(y)\Delta\Phi_{\left(  \xi\right)  }\ ,\ \ \ g(y)=\left[  1-\exp(y)\right]
/y\ ,\ \ \ y(\Gamma|-\Delta\Phi_{\left(  \xi\right)  })=-\frac{1}{4i\hbar
}\varepsilon_{ab}\int dt\left\{  \left\{  \Delta\Phi_{\left(  \xi\right)
},\Omega^{a}\right\}  ,\Omega^{b}\right\}  \ . \label{LambYM}%
\end{equation}
According to (\ref{where}), we have%
\begin{equation}
\frac{1}{2}\varepsilon_{ab}\left\{  \left\{  \Delta\Phi_{\left(  \xi\right)
},\Omega^{a}\right\}  ,\Omega^{b}\right\}  =-\frac{\Delta\xi}{2}\int
d\boldsymbol{x}\left(  \pi^{\mathsf{u}}\pi^{\mathsf{u}}-\frac{1}%
{24}f^{\mathsf{vuw}}f^{\mathsf{wts}}C^{\mathsf{s}a}C^{\mathsf{t}%
c}C^{\mathsf{u}b}C^{\mathsf{v}d}\varepsilon_{ab}\varepsilon_{cd}\right)  \ ,
\label{auxYM}%
\end{equation}
which implies
\begin{equation}
y(\Gamma|-\Delta\Phi_{\left(  \xi\right)  })=\frac{\Delta\xi}{2i\hbar}\int
d^{D}x\left(  \pi^{\mathsf{u}}\pi^{\mathsf{u}}-\frac{1}{24}f^{\mathsf{vuw}%
}f^{\mathsf{wts}}C^{\mathsf{s}a}C^{\mathsf{t}c}C^{\mathsf{u}b}C^{\mathsf{v}%
d}\varepsilon_{ab}\varepsilon_{cd}\right)  \ , \label{yYM}%
\end{equation}
and, due to (\ref{funcdeplafin}), (\ref{4.9}), (\ref{DFxi}), the corresponding
parameters $\lambda_{a}(\Gamma|-\Delta\Phi_{\left(  \xi\right)  })$ have the
form%
\begin{equation}
\lambda_{a}(\Gamma|-\Delta\Phi_{\left(  \xi\right)  })=-\frac{1}{2i\hbar
}\varepsilon_{ab}g(y)\int dt\ \left\{  \Delta\Phi_{\left(  \xi\right)
},\Omega^{b}\right\}  =\frac{\Delta\xi}{4i\hbar}\varepsilon_{ab}g(y)\int
d^{D}x\ \pi^{\mathsf{u}}C^{\mathsf{u}b} \label{lamaxi}%
\end{equation}
and generate the transition from an $R_{\xi}$-like gauge to another $R_{\xi}%
$-like gauge corresponding to $\xi+\Delta\xi$.

For comparison, notice that in the Lagrangian approach of \cite{MRnew} the
transition from an $R_{\xi}$-like gauge to an $R_{\xi+\Delta\xi}$-like gauge
is described by the finite BRST-antiBRST transformation%
\begin{align}
\Delta A_{\mu}^{\mathsf{m}}  &  =D_{\mu}^{\mathsf{mn}}C^{\mathsf{n}a}%
\lambda_{a}-\frac{1}{2}\left(  D_{\mu}^{\mathsf{mn}}B^{\mathsf{n}}+\frac{1}%
{2}f^{\mathsf{mnl}}C^{\mathsf{l}a}D_{\mu}^{\mathsf{nk}}C^{\mathsf{k}%
b}\varepsilon_{ba}\right)  \lambda^{2}\ ,\label{DAmm}\\
\Delta B^{\mathsf{m}}  &  =-\frac{1}{2}\left(  f^{\mathsf{mnl}}B^{\mathsf{l}%
}C^{\mathsf{n}a}+\frac{1}{6}f^{\mathsf{mnl}}f^{\mathsf{lrs}}C^{\mathsf{s}%
b}C^{\mathsf{r}a}C^{\mathsf{n}c}\varepsilon_{cb}\right)  \lambda
_{a}\ ,\label{DBm}\\
\Delta C^{\mathsf{m}a}  &  =\left(  \varepsilon^{ab}B^{\mathsf{m}}-\frac{1}%
{2}f^{\mathsf{mnl}}C^{\mathsf{l}a}C^{\mathsf{n}b}\right)  \lambda_{b}-\frac
{1}{2}\left(  f^{\mathsf{mnl}}B^{\mathsf{l}}C^{\mathsf{n}a}+\frac{1}%
{6}f^{\mathsf{mnl}}f^{\mathsf{lrs}}C^{\mathsf{s}b}C^{\mathsf{r}a}%
C^{\mathsf{n}c}\varepsilon_{cb}\right)  \lambda^{2}\ , \label{DCma}%
\end{align}
with the field-dependent parameters $\lambda_{a}=\lambda_{a}\left(
\phi\right)  $%
\begin{align}
\lambda_{a}  &  =\frac{\Delta\xi}{4i\hbar}\varepsilon_{ab}\int d^{D}x\ \left(
B^{\mathsf{n}}C^{\mathsf{n}b}+\frac{1}{2}f^{\mathsf{nml}}C^{\mathsf{l}%
c}C^{\mathsf{m}b}C^{\mathsf{n}d}\varepsilon_{cd}\right) \nonumber\\
&  \times\sum_{n=0}^{\infty}\frac{1}{\left(  n+1\right)  !}\left[  \frac
{1}{4i\hbar}\Delta\xi\int d^{D}y\ \left(  B^{\mathsf{u}}B^{\mathsf{u}}%
-\frac{1}{24}\ f^{\mathsf{uwt}}f^{\mathsf{trs}}C^{\mathsf{s}e}C^{\mathsf{r}%
p}C^{\mathsf{w}g}C^{\mathsf{u}q}\varepsilon_{eg}\varepsilon_{pq}\right)
\right]  ^{n}\ . \label{lamaxi2}%
\end{align}

Concluding, note that a finite change $\Phi\rightarrow\Phi+\Delta\Phi$ of the
gauge condition induces a finite change of a function $\mathcal{G}_{\Phi
}(\Gamma)$ or a functional ${G}_{\Phi}(\Gamma)$, so that in the reference
frame corresponding to the gauge $\Phi+\Delta\Phi$ it can be represented,
according to (\ref{DelFfrule}), (\ref{funcdeplafin}), as follows:%
\begin{equation}
\mathcal{G}_{\Phi+\Delta\Phi}=\mathcal{G}_{\Phi}+\left(  s^{a}\mathcal{G}%
_{\Phi}\right)  \lambda_{a}\left(  \Delta\Phi\right)  +\frac{1}{4}\left(
s^{2}\mathcal{G}_{\Phi}\right)  \lambda_{a}\left(  \Delta\Phi\right)
\lambda^{a}\left(  \Delta\Phi\right)  \,,\label{finvarfun}%
\end{equation}
which is an extension of the infinitesimal change $\mathcal{\mathcal{G}}%
_{\Phi}\rightarrow\mathcal{G}_{\Phi}+\delta\mathcal{G}_{\Phi}$ induced by a
variation of the gauge, $\Phi\rightarrow\Phi+\delta\Phi$,
\begin{equation}
\mathcal{G}_{\Phi+\delta\Phi}=\mathcal{G}_{\Phi}-\frac{i}{2\hbar}\left(
s^{a}\mathcal{G}_{\Phi}\right)  \left(  \int dts_{a}\delta\Phi(t)\right)
\,,\label{infinvarfungt}%
\end{equation}
corresponding, in the particular case $\mathcal{G}_{\Phi}\left(  \eta\right)
$, to the gauge transformations%
\begin{equation}
\delta\eta=\{\eta,T_{\alpha_{0}}\}C^{\alpha_{0}{}a}\int dt(s_{a}\delta
\Phi)(t)\equiv\{\eta,T_{\alpha_{0}}\}\zeta^{\alpha_{0}}\ ,\ \ \ \mathrm{for}%
\ \ \ \zeta^{\alpha_{0}}=C^{\alpha_{0}{}a}\int dt(s_{a}\delta\Phi
)(t)\,,\label{infinvarfungt2}%
\end{equation}
which in Yang--Mills theories are given by functions $\zeta^{\mathsf{u}%
}(t,\boldsymbol{x})$:%
\begin{equation}
\delta\mathcal{G}_{\Phi}=\mathcal{G}_{\Phi+\delta\Phi}-\mathcal{G}_{\Phi}=\int
d\boldsymbol{x}\,\frac{\delta\mathcal{G}_{\Phi}(t)}{\delta\eta(t,\boldsymbol{x}%
)}\left\{  \eta(t,\boldsymbol{x}),T^{\mathsf{u}}(t,\boldsymbol{x})\right\}
\zeta^{\mathsf{u}}(t,\boldsymbol{x})\,,\ \ \ \mathrm{where}\ \ \ \zeta
^{\mathsf{u}}\left(  t,\boldsymbol{x}\right)  =-\frac{i}{2\hbar}%
C^{\mathsf{u}a}\left(  t,\boldsymbol{x}\right)  \int dt^{\prime}\,{\left(
s_{a}\delta\Phi\right)  }(t^{\prime})\,.\label{infinvarfun}%
\end{equation}
Due to the presence of the term with $s^{2}\mathcal{G}_{\Phi}$ in the finite
gauge variation of a function $\mathcal{G}_{\Phi}(\eta)$, depending on the
classical phase-space coordinates $\eta$, the representation (\ref{finvarfun})
is more general than that which would correspond to the generalized
Hamiltonian scheme \cite{BRST3, Henneaux1}, having a form similar to
(\ref{infinvarfun}), and therefore also to (\ref{infinvarfungt}).

\section{Conclusion}

\label{Concl} In the present work, we have proposed the concept of
finite BRST-antiBRST transformations for phase-space variables and
trajectories in the $\mathrm{Sp}(2)$-covariant generalized
Hamiltonian quantization \cite{BLT1h, BLT2h}. This concept
is realized in the form (\ref{deffin1}),
(\ref{deffin_traj}), being polynomial in powers of a constant $\mathrm{Sp}%
\left(  2\right)  $-doublet of anticommuting Grassmann parameters $\lambda
_{a}$ and leaving the integrand in the partition function for dynamical
systems subject to first-class constraints invariant to all orders of the
constant doublet $\lambda_{a}$. We have established the fact that the finite
BRST-antiBRST transformations with a constant doublet $\lambda_{a}$ are
canonical transformations.

We have introduced finite field-dependent BRST-antiBRST transformations as
polynomials in powers of the $\mathrm{Sp}\left(  2\right)  $-doublet of
Grassmann-odd functionals $\lambda_{a}(\Gamma)$, depending on the entire set
of phase-space variables for an arbitrary constrained dynamical system in the
$\mathrm{Sp}(2)$-covariant generalized Hamiltonian quantization. In a special
case of functionally-dependent $\lambda_{a}$, we have obtained modified Ward
identities (\ref{mWI}), depending on $\lambda_{a}$, and therefore also on a
variation of the gauge Boson, which leads to Ward identities for Green's
functions with an additional weight function constructed from $\lambda_{a}$,
and allows one to study the problem of gauge dependence (\ref{GDI}) and to
obtain the standard Ward identities with constant $\lambda_{a}$. We have
calculated the Jacobian (\ref{superJaux}), (\ref{superJ1}) corresponding to
this change of variables, by using a special class of transformations with
functionally-dependent parameters $\lambda_{a}(\Gamma)=\int dt\ s_{a}%
\Lambda(\Gamma)$ for a Grassmann-even function $\Lambda(\phi)$ and
Grassmann-odd generators $s_{a}$ of BRST-antiBRST transformations in
Hamiltonian formalism.

In comparison with finite field-dependent BRST--BFV
transformations \cite{BLThf} in the generalized Hamiltonian
formalism \cite{BFV,Henneaux1}, where a change of the gauge
corresponds to a unique (up to BRST-exact terms) field-dependent
parameter, it is only functionally-dependent finite BRST-antiBRST
transformations with $\lambda_{a}=\int dt s_{a}\Lambda
(\Gamma(t)|\Delta\Phi)$ that are in one-to-one correspondence with
$\Delta \Phi$. We have found in (\ref{solcompeq2}) a solution
$\Lambda(\Delta\Phi)$ to the compensation equation (\ref{eqexpl})
for an unknown function $\Lambda$ generating an $\mathrm{Sp}\left(
2\right)  $-doublet $\lambda_{a}$ in (\ref{funcdeplafin}), in
order to establish a relation between the partition functions
$Z_{\Phi}$ and $Z_{\Phi+\Delta\Phi}$, with the respective
action $S_{H,\Phi}$ in a certain gauge induced by a gauge Boson
$\Phi$ and the action $S_{H,\Phi+\Delta\Phi}$ induced by a
different gauge $\Phi+\Delta\Phi$. This makes it possible to
investigate the problem of gauge-dependence for the generating
functional $Z_{\Phi}(I)$ under a finite change of the gauge in the
form (\ref{GDI}), leading to the gauge-independence of the
physical $S$-matrix.

In terms of the potential $\Lambda$ which generates finite field-dependent
BRST-antiBRST transformations, we have explicitly constructed (\ref{lamaxi})
the parameters $\lambda_{a}$ generating a change of the gauge in the path
integral for Yang--Mills theories within a class of linear $R_{\xi}$-like
gauges in Hamiltonian formalism, related to even-valued gauge-fixing functions
$\Phi_{\left(  \xi\right)  }$, with $\xi=0,1$ corresponding to the respective
Landau and Feynman (covariant) gauges in Hamiltonian formalism. We have
established, after integrating over momenta in the Hamiltonian path integral
for an arbitrary gauge Boson $\Phi_{\left(  \xi\right)  }$, that the result
(\ref{S(A,B,C)}) is identical with\ the generating functional of Green's
functions\ recently obtained in \cite{MRnew} by the Lagrangian BRST-antiBRST
quantization of the Yang--Mills theory, which justifies the unitarity of the
$S$-matrix in the Lagrangian approach of \cite{MRnew}. We have suggested an
explicit rule (\ref{finvarfun}) of calculating the value of an arbitrary
function $\mathcal{G}_{\Phi}(\Gamma)$ given in a certain gauge induced by the
Bosonic function $\Phi$, by using any other gauge $\Phi+\Delta\Phi$ in terms
of finite field-dependent BRST-antiBRST transformations with
functionally-dependent parameters $\lambda_{a}\left(  \Delta\Phi\right)  $ in
(\ref{funcdeplafin}), constructed using a finite variation $\Delta\Phi$.

Notice that, upon submission of this work to arXiv, we became
aware of the article \cite{BLThfext}, in which similar problems
are discussed. As compared to our present work, the study of
\cite{BLThfext} deals with a calculation of the Jacobian for a
change of variables given by BRST-antiBRST (BRST--BFV by the
terminology of \cite{BLThfext}) transformations with functionally
independent field-dependent odd-valued parameters
$\lambda_a(\Gamma)$, subsequently used to formulate a compensation
equation, similar to (\ref{eqexpl}), but having a $2\times 2$
matrix form, which satisfies the condition of resolvability only
for functionally-dependent parameters, $\lambda_{a}=\int dt\,
s_{a}\Lambda (\Gamma(t)|\Delta\Phi)$, whose form was first
announced in our work \cite{MRnew}.

There are various directions for extending the results of the
present work: the study of soft BRST--BFV and BRST-antiBRST
symmetry breaking in the respective generalized Hamiltonian
formulations \cite{BRST3, Henneaux1} and \cite{BLT1h, BLT2h}; the
study of the Gribov problem \cite{Gribov} in the BRST--BFV and
BRST-antiBRST generalized Hamiltonian formulations and its
relation to the Lagrangian description \cite{MRnew,LL2};
the calculation of Jacobians corresponding to BRST-antiBRST
transformations linear in finite field-dependent parameters, as
well as transformations with polynomial (group-like) but not
functionally-dependent parameters $\lambda_a$ [leading to an
essentially different representation for the Jacobian than the one
in (\ref{superJ1})], which is a substantial part of our current
study \cite{MRmew4}. The other problems from the above list are
also planned to be examined in our forthcoming works.

\section*{Acknowledgments}

The authors are grateful to R. Metsaev for useful remarks and to
the participants of the International Conference QFTG'2014, Tomsk,
July 28--August 3, 2014. The study was supported by the RFBR grant
under Project No. 12-02-00121 and by the grant of Leading
Scientific Schools of the Russian Federation under Project No.
88.2014.2. The work was also partially supported by the Ministry
of Science of the Russian Federation, Grant No. 2014/223.

\appendix

\section*{Appendix}

\section{Calculation of Jacobians}

\label{AppA} \renewcommand{\theequation}{\Alph{section}.\arabic{equation}} \setcounter{equation}{0}

In this Appendix, we present the calculation of the Jacobian (\ref{measure}),
(\ref{superJ}) induced in the functional integral\ (\ref{ZPI}) by finite
BRST-antiBRST transformations of phase-space trajectories (\ref{deffin_traj})
with an $\mathrm{Sp}\left(  2\right)  $-doublet $\lambda_{a}$ of anticommuting
parameters, considered in the case $\lambda_{a}=\mathrm{const}$ and in the
case of functionals $\lambda_{a}\left(  \Gamma\right)  $ of a special form,
$\lambda_{a}\left(  \Gamma\right)  =\int dt\ s_{a}\Lambda\left(
\Gamma\right)  $. To this end, let us choose the parameters of
(\ref{deffin_traj}) in the most general form $\lambda_{a}=\lambda_{a}\left(
\Gamma\right)  $ and consider the even matrix $M$ in (\ref{superJ}) with the
elements $M_{q}^{p}\left(  t^{\prime}|t^{\prime\prime}\right)  \equiv
M_{q|t^{\prime},t^{\prime\prime}}^{p}$, $\varepsilon(M_{q|t^{\prime}%
,t^{\prime\prime}}^{p})=\varepsilon_{p}+\varepsilon_{q}$,%
\begin{align}
&  M_{q|t^{\prime},t^{\prime\prime}}^{p}=\frac{\delta\left(  \Delta
\Gamma_{t^{\prime}}^{p}\right)  }{\delta\Gamma_{t^{\prime\prime}}^{q}%
}=U_{q|t^{\prime},t^{\prime\prime}}^{p}+V_{q|t^{\prime},t^{\prime\prime}}%
^{p}+W_{q|t^{\prime},t^{\prime\prime}}^{p}\ ,\ \ V_{q|t^{\prime}%
,t^{\prime\prime}}^{p}=\left(  V_{1}\right)  _{q|t^{\prime},t^{\prime\prime}%
}^{p}+\left(  V_{2}\right)  _{q|t^{\prime},t^{\prime\prime}}^{p}\ ,\label{M}\\
&  U_{q|t^{\prime},t^{\prime\prime}}^{p}=X_{t^{\prime}}^{pa}\frac
{\delta\lambda_{a}}{\delta\Gamma_{t^{\prime\prime}}^{q}}\ ,\ \ \left(
V_{1}\right)  _{q|t^{\prime},t^{\prime\prime}}^{p}=\lambda_{a}\frac{\delta
X_{t^{\prime}}^{pa}}{\delta\Gamma_{t^{\prime\prime}}^{q}}\left(  -1\right)
^{\varepsilon_{p}+1}\ ,\ \ \left(  V_{2}\right)  _{q|t^{\prime},t^{\prime
\prime}}^{p}=\lambda_{a}Y_{t^{\prime}}^{p}\frac{\delta\lambda^{a}}%
{\delta\Gamma_{t^{\prime\prime}}^{q}}\left(  -1\right)  ^{\varepsilon_{p}%
+1}\ ,\ \ W_{q|t^{\prime},t^{\prime\prime}}^{p}=-\frac{1}{2}\lambda^{2}%
\frac{\delta Y_{t^{\prime}}^{p}}{\delta\Gamma_{t^{\prime\prime}}^{q}%
}\ ,\nonumber
\end{align}
where the functions $X_{t}^{pa}=X^{pa}\left(  \Gamma\left(  t\right)  \right)
$ and $Y_{t}^{p}=Y^{p}\left(  \Gamma\left(  t\right)  \right)  $ are given by%
\begin{equation}
X_{t}^{pa}=\left(  s^{a}\Gamma^{pa}\right)  _{t}\ ,\ \ \ Y_{t}^{p}=-\frac
{1}{2}\left(  s^{2}\Gamma^{p}\right)  _{t}=-\frac{1}{2}\varepsilon_{ab}\int
dt^{\prime}\frac{\delta X_{t}^{pa}}{\delta\Gamma_{t^{\prime}}^{q}}%
X_{t^{\prime}}^{Bb} \label{defXY}%
\end{equation}
and possess the properties%
\begin{equation}
\int dt^{\prime}\ \frac{\delta X_{t}^{pa}}{\delta\Gamma_{t^{\prime}}^{q}%
}X_{t^{\prime}}^{qb}=\varepsilon^{ab}Y_{t}^{p}\ ,\ \ \ \int dt^{\prime}%
\ \frac{\delta Y_{t}^{p}}{\delta\Gamma_{t^{\prime}}^{q}}X_{t^{\prime}}%
^{qa}=0\ ,\ \ \ \int dt\ \frac{\delta X_{t}^{pa}}{\delta\Gamma_{t}^{p}}=0\ .
\label{propXY}%
\end{equation}
Indeed, due to the anticommutativity, $s^{a}s^{b}+s^{b}s^{a}=0$, and
nilpotency, $s^{a}s^{b}s^{c}=0$, of the generators $s^{a}$, we have%
\begin{align}
\int dt^{\prime}\frac{\delta X_{t}^{pa}}{\delta\Gamma_{t^{\prime}}^{q}%
}X_{t^{\prime}}^{qb}  &  =\int dt^{\prime}\frac{\delta X_{t}^{pa}}%
{\delta\Gamma_{t^{\prime}}^{q}}\left(  s^{b}\Gamma^{q}\right)  _{t^{\prime}%
}=\left(  s^{a}s^{b}\Gamma^{p}\right)  _{t}=\varepsilon^{ab}Y_{t}%
^{p}\ ,\label{id0}\\
\int dt^{\prime}\ \frac{\delta Y_{t}^{p}}{\delta\Gamma_{t^{\prime}}^{q}%
}X_{t^{\prime}}^{qa}  &  =\int dt^{\prime}\ \frac{\delta Y_{t}^{p}}%
{\delta\Gamma_{t^{\prime}}^{q}}\left(  s^{a}\Gamma^{q}\right)  _{t^{\prime}%
}=\left(  s^{a}Y^{p}\right)  _{t}=-\frac{1}{2}\varepsilon_{bc}s^{a}\left(
s^{b}s^{c}\Gamma^{p}\right)  _{t}=0\ ; \label{id1}%
\end{align}
besides, we have%
\begin{align}
&  X_{t}^{pa}=\left\{  \Gamma^{p},\Omega^{a}\right\}  _{t}\ ,\ \ \ \Gamma
^{p}=\left(  P_{A},Q^{A}\right)  \ ,\nonumber\\
&  X_{A|t}^{a}=\left\{  P_{A},\Omega^{a}\right\}  _{t}=\left(  -1\right)
^{\varepsilon_{A}+1}\left.  \frac{\partial\Omega^{a}}{\partial Q^{A}%
}\right\vert _{t}\ ,\ \ \ X_{t}^{Aa}=\left\{  Q^{A},\Omega^{a}\right\}
_{t}=\left.  \frac{\partial\Omega^{a}}{\partial P_{A}}\right\vert
_{t}\ ,\nonumber\\
&  \int dt\ \frac{\delta X_{t}^{pa}}{\delta\Gamma_{t}^{p}}=\int dt\ \left[
\frac{\delta X_{A}^{a}\left(  t\right)  }{\delta P_{A}\left(  t\right)
}+\frac{\delta X^{Aa}\left(  t\right)  }{\delta Q^{A}\left(  t\right)
}\right]  =\delta\left(  0\right)  \int dt\ \left[  \left(  -1\right)
^{\varepsilon_{A}+1}\frac{\partial}{\partial P_{A}}\left(  \frac
{\partial\Omega^{a}}{\partial Q^{A}}\right)  +\frac{\partial}{\partial Q^{A}%
}\left(  \frac{\partial\Omega^{a}}{\partial P_{A}}\right)  \right]
_{t}\nonumber\\
&  =\delta\left(  0\right)  \int dt\ \left[  -\frac{\partial}{\partial Q^{A}%
}\left(  \frac{\partial\Omega^{a}}{\partial P_{A}}\right)  +\frac{\partial
}{\partial Q^{A}}\left(  \frac{\partial\Omega^{a}}{\partial P_{A}}\right)
\right]  _{t}\equiv0\ . \label{id2}%
\end{align}

Recall that the Jacobian $\exp\left(  \Im\right)  $ induced by the finite
BRST-antiBRST transformation (\ref{deffin_traj}) with the corresponding matrix
$M$ in (\ref{M}) is given by (\ref{superJ}), namely,%
\begin{equation}
\Im=\mathrm{Str}\ln\left(  \mathbb{I}+M\right)  =-\sum_{n=1}^{\infty}%
\frac{\left(  -1\right)  ^{n}}{n}\,\,\mathrm{Str}\left(  M^{n}\right)  \,.
\label{Mstruct}%
\end{equation}
In order to calculate the Jacobian explicitly in the cases $\lambda
_{a}=\mathrm{const}$ and $\lambda_{a}=\int dt\ s_{a}\Lambda$, it is sufficient
to use the above properties (\ref{propXY}), the identities $\lambda_{a}%
\lambda^{2}=\lambda^{4}\equiv0$, the definitions%
\begin{equation}
\left(  AB\right)  _{q|t^{\prime},t^{\prime\prime}}^{p}=\int dt\ \left(
A\right)  _{r|t^{\prime},t}^{p}\left(  B\right)  _{q|t,t^{\prime\prime}}%
^{r}\ ,\ \ \ \mathrm{Str}\left(  A\right)  =\left(  -1\right)  ^{\varepsilon
_{p}} \int dt\ \left(  A\right)  _{p|t,t}^{p} \label{ABstruct}%
\end{equation}
and the property of supertrace%
\[
\mathrm{Str}\left(  AB\right)  =\mathrm{Str}\left(  BA\right)  \,,
\]
\hspace{-1ex}{\begin{table}{{
\begin{center}
\begin{tabular}{||l|l||}\hline\hline
{Hamiltonian\ formalism} & {Lagrangian\ formalism}
 \\
\hline\hline $ \Gamma_{t}^{p}\ ,\ \Delta\Gamma_{t}^{p}=\left(
s^{a}\Gamma_{t}^{p}\right) \lambda_{a}+\frac{1}{4}\left(
s^{2}\Gamma_{t}^{p}\right)  \lambda^{2}$ &  $\phi^{A}\ ,\
\Delta\phi^{A}=\left(  s^{a}\phi^{A}\right)  \lambda_{a}+\frac
{1}{4}\left(  s^{2}\phi^{A}\right)  \lambda^{2}\ ,\ A=\left(  p,t\right)$\\
 $ \frac{\delta\left(  \Delta\Gamma_{t^{\prime}}^{p}\right)  }{\delta
\Gamma_{t^{\prime\prime}}^{q}}=M_{q|t^{\prime},t^{\prime\prime}}^{p}
$
    & $\frac{\delta\left(  \Delta\phi^{A}\right)  }{\delta\phi^{B}}=M_{B}%
^{A}\ ,\ A=\left(  p,t^{\prime}\right)  \ ,\ B=\left(
q,t^{\prime\prime
}\right)$  \\
 $s^{a}\Gamma_{t}^{p}=X_{t}^{pa},\ Y_{t}^{p}=-\frac{1}{2}\varepsilon_{ab}\int
dt^{\prime}\frac{\delta X_{t}^{pa}}{\delta\Gamma_{t^{\prime}}^{q}}%
X_{t^{\prime}}^{Bb}$
    &  $ s^{a}\phi^{A}=X^{Aa},\ Y^{A}=-\frac{1}{2}%
\varepsilon_{ab}\frac{\delta X^{Aa}}{\delta\phi^{B}}X^{Bb}$ \\
$ \int dt^{\prime}\ \frac{\delta X_{t}^{pa}}{\delta\Gamma_{t^{\prime}}^{q}%
}X_{t^{\prime}}^{qb}=\varepsilon^{ab}Y_{t}^{p}\ ,\ \int dt^{\prime}%
\ \frac{\delta Y_{t}^{p}}{\delta\Gamma_{t^{\prime}}^{q}}X_{t^{\prime}}%
^{qa}=\int dt\ \frac{\delta X_{t}^{pa}}{\delta\Gamma_{t}^{p}}=0$
   & $  \frac{\delta X^{Aa}}{\delta\phi^{B}}X^{Bb}=\varepsilon^{ab}Y^{A}%
\ ,\ \frac{\delta Y^{A}}{\delta\phi^{B}}X^{Bb}=\frac{\delta
X^{Aa}}{\delta
\phi^{A}}=0 $  \\
 $ M_{q|t^{\prime},t^{\prime\prime}}^{p}=U_{q|t^{\prime},t^{\prime\prime}}%
^{p}+V_{q|t^{\prime},t^{\prime\prime}}^{p}+W_{q|t^{\prime},t^{\prime\prime}%
}^{p} $ &
   $  M_{B}^{A}=P_{B}^{A}+Q_{B}^{A}+R_{B}^{A} $   \\
 $ V_{q|t^{\prime},t^{\prime\prime}}^{p}=\left(  V_{1}\right)  _{q|t^{\prime
},t^{\prime\prime}}^{p}+\left(  V_{2}\right)  _{q|t^{\prime},t^{\prime\prime}%
}^{p}$
    &  $  Q_{B}^{A}=\left(  Q_{1}\right)  _{B}^{A}+\left(  Q_{2}\right)
_{B}^{A}$  \\
 $ \left(  V_{1}\right)  _{q|t^{\prime},t^{\prime\prime}}^{p}=\lambda_{a}%
\frac{\delta
X_{t^{\prime}}^{pa}}{\delta\Gamma_{t^{\prime\prime}}^{q}}\left(
-1\right)  ^{\varepsilon_{p}+1}  $
    & $\left(  Q_{1}\right)  _{B}^{A}%
=\lambda_{a}\frac{\delta X^{Aa}}{\delta\phi^{B}}\left(  -1\right)
^{\varepsilon_{A}+1} $  \\
$  \left(  V_{2}\right)
_{q|t^{\prime},t^{\prime\prime}}^{p}=\lambda
_{a}Y_{t^{\prime}}^{p}\frac{\delta\lambda^{a}}{\delta\Gamma_{t^{\prime\prime}%
}^{q}}\left(  -1\right)  ^{\varepsilon_{p}+1} $ & $ \left(
Q_{2}\right)
_{B}^{A}=\lambda_{a}Y^{A}\frac{\delta\lambda^{a}}{\delta\phi^{B}}\left(
-1\right)  ^{\varepsilon_{A}+1}$ \\
$U_{q|t^{\prime},t^{\prime\prime}}^{p}=X_{t^{\prime}}^{pa}\frac
{\delta\lambda_{a}}{\delta\Gamma_{t^{\prime\prime}}^{q}}\ ,\
W_{q|t^{\prime
},t^{\prime\prime}}^{p}=-\frac{1}{2}\lambda^{2}\frac{\delta Y_{t^{\prime}}%
^{p}}{\delta\Gamma_{t^{\prime\prime}}^{q}} $ & $
P_{B}^{A}=X^{Aa}\frac
{\delta\lambda_{a}}{\delta\phi^{B}}\ ,\ R_{B}^{A}=-\frac{1}{2}\lambda^{2}%
\frac{\delta Y^{A}}{\delta\phi^{B}}$ \\
$ \mathrm{Str}\left(  V_{1}\right)  =\mathrm{Str}\left(  UW\right)
=0\ ,\ \mathrm{Str}\left(  V_{1}^{2}\right)  =2\mathrm{Str}\left(
W\right) $ & $  \mathrm{Str}\left(  Q_{1}\right)
=\mathrm{Str}\left(  PR\right)
=0\ ,\ \mathrm{Str}\left(  Q_{1}^{2}\right)  =2\mathrm{Str}\left(  R\right)$ \\
$\lambda_{a}=\mathrm{const}:\mathrm{\ }U=V_{2}=0\ ,\ \Im=0 $ & $
\lambda
_{a}=\mathrm{const}:\mathrm{\ }P=Q_{2}=0\ ,\ \Im=0$ \\
$\lambda_{a}=\int dt\ s_{a}\Lambda\left(  \Gamma\left(  t\right)
\right)  :
  $ & $ \lambda_{a}=s_{a}\Lambda\left(  \phi\right): $ \\
$ U^{2}=f\cdot U\ ,\ VU=\left(  1+f\right)  \cdot V_{2}\ ,\ \ f=-\frac{1}%
{2}\mathrm{Str}\left(  U\right) $ & $ P^{2}=f\cdot P\ ,\ QP=\left(
1+f\right)  \cdot Q_{2}\ ,\ \ f=-\frac{1}{2}\mathrm{Str}\left(  P\right)$ \\
$ \int dt\ \frac{\delta\lambda_{b}}{\delta\Gamma_{t}^{p}}X_{t}^{pa}%
=s^{a}\lambda_{b}=\delta_{b}^{a}f\ ,\ \ f=\frac{1}{2}s^{a}\lambda_{a}%
=-\frac{1}{2}\int dt\ \left(  s^{2}\Lambda\right)  \left(  t\right)  $ & $ \frac{\delta\lambda_{b}}{\delta\phi^{A}}X^{Aa}=s^{a}\lambda_{b}=\delta_{b}%
^{a}f\ ,\ \ f=\frac{1}{2}s^{a}\lambda_{a}=-\frac{1}{2}s^{2}\Lambda$ \\
 $ \Im=-2\mathrm{\ln}\left(  1+f\right) $
    & $  \Im=-2\mathrm{\ln}\left(
1+f\right)$ \\
   \hline\hline
\end{tabular}
\end{center}}} \vspace{-2ex}\caption{Correspondence of the  matrix elements in Lagrangian and
Hamiltonian formalisms.\label{table in}}\end{table} which takes
place for any even matrices $A$, $B$. In this setting, the task of
calculation is formally identical with the one carried out in our
previous work \cite{MRnew} that deals with the calculation of
Jacobians induced by finite BRST-antiBRST transformations in the
Lagrangian approach to the Yang--Mills type of theories. Since the
corresponding reasonings and results of \cite{MRnew} in the
Lagrangian formalism can be literally reproduced in the
Hamiltonian formalism of the present work, we
give them briefly in Table~\ref{table in}.%

Therefore, the Jacobians $\exp\left(  \Im\right)  $ corresponding to the cases
$\lambda_{a}=\mathrm{const\ }$and $\lambda_{a}=\int dt\ s_{a}\Lambda\left(
\Gamma\left(  t\right)  \right)  $ are given by%
\begin{align}
&  \lambda_{a}=\mathrm{const:\ \ \ }\Im=0\ ,\label{const}\\
&  \lambda_{a}\left(  \Gamma\right)  =\int dt\ s_{a}\Lambda\left(
\Gamma\left(  t\right)  \right)  :\ \ \ \Im=-2\mathrm{\ln}\left(  1+f\right)
\ ,\ \ \ \ f=-\frac{1}{2}\int dt\ \left(  s^{2}\Lambda\right)  _{t}\ .
\label{potent}%
\end{align}

\end{document}